\documentclass[aps,preprint,letterpaper,showpacs,prd]{revtex4-1}%
\usepackage{amsfonts}
\usepackage{amsmath}
\usepackage{amssymb}
\usepackage{graphicx}%
\usepackage[pdftex,colorlinks=true,linkcolor=blue,citecolor=blue,filecolor=blue,urlcolor=black
]{hyperref}
\providecommand{\U}[1]{\protect\rule{.1in}{.1in}}
\begin{document}
\title{Black hole for the Einstein-Chern-Simons gravity}
\author{C.A.C. Quinzacara}
\email[Electronic address: ]{crcortes@udec.cl}
\author{P. Salgado}
\email[Electronic address: ]{pasalgad@udec.cl}
\affiliation{Departamento de F\'{\i}sica, Universidad de Concepci\'{o}n, Casilla 160-C, Concepci\'{o}n, Chile}

\begin{abstract}
We consider a 5-dimensional action which is composed of a gravitational
sector and a sector of matter, where the gravitational sector is given by a
Einstein-Chern-Simons gravity action instead of the Einstein-Hilbert action.

We obtain the Einstein-Chern-Simons $(EChS)$ field equations together with
its spherically symmetric solution, which lead, in certain limit, to the
standard five dimensional solution of the Einstein-Cartan field equations.

It is found the conditions under which the $EChS$ field equations admits
black hole type solutions. The maximal extension and conformal
compactification are also studied
\end{abstract}

\maketitle

\flushbottom

\section{Introduction}
According to the principles of general relativity (GR), the spacetime is a
dynamical object which has independent degrees of freedom, and is governed
by dynamical equations, namely the Einstein field equations. This means that
in GR the geometry is dynamically determined. Therefore, the construction of
a gauge theory of gravity requires an action that does not consider a fixed
space-time background. An five dimensional action for gravity fulfilling
these conditions is the five-dimensional Chern--Simons AdS gravity action,
which can be written as

\begin{equation}
L_{\mathrm{AdS}}^{\left( 5\right) }=\kappa \left( \frac{1}{5l^{5}}\epsilon
_{a_{1}\cdots a_{5}}e^{a_{1}}\cdots e^{a_{5}}+\frac{2}{3l^{3}}\epsilon
_{a_{1}\cdots a_{5}}R^{a_{1}a_{2}}e^{a_{3}}\cdots e^{a_{5}}+\frac{1}{l}%
\epsilon _{a_{1}\cdots a_{5}}R^{a_{1}a_{2}}R^{a_{3}a_{4}}e^{a_{5}}\right) ,
\end{equation}%
where $e^{a}$ corresponds to the 1-form \emph{vielbein}, and $R^{ab}=d\omega
^{ab}+\omega _{\text{ }c}^{a}\omega ^{cb}$ to the Riemann curvature in the
first order formalism \cite{champ1}, \cite{champ2}, \cite{zan1}.

If Chern-Simons theories are the appropriate gauge-theories to provide a
framework for the gravitational interaction, then these theories must
satisfy the correspondence principle, namely they must be related to General
Relativity.

In ref. \cite{salg1} was recently shown that the standard, five-dimensional
General Relativity (without a cosmological constant) can be obtained from
Chern-Simons gravity theory for a certain Lie algebra $\mathcal{B}$. The
Chern-Simons Lagrangian is built from a $\mathcal{B}$-valued, one-form gauge
connection $A$ which depends on a scale parameter $l$ which can be
interpreted as a coupling constant that characterizes different regimes
within the theory. The $\mathcal{B}$ algebra, on the other hand, is obtained
from the $AdS$ algebra and a particular semigroup $S$ by means of the
S-expansion procedure introduced in refs. \cite{salg2}, \cite{salg3}. The
field content induced by $\mathcal{B}$ includes the vielbein $e^{a}$, the
spin connection $\omega^{ab}$ and two extra bosonic fields $h^{a}$ and $%
k^{ab}$.

The five dimensional Chern-Simons Lagrangian for the $\mathcal{B}$ algebra
is given by \cite{salg1}:

\begin{equation}
L_{\mathrm{ChS}}^{(5)}=\alpha _{1}l^{2}\varepsilon
_{abcde}R^{ab}R^{cd}e^{e}+\alpha _{3}\varepsilon _{abcde}\left( \frac{2}{3}%
R^{ab}e^{c}e^{d}e^{e}+2l^{2}k^{ab}R^{cd}T^{\text{ }e}+l^{2}R^{ab}R^{cd}h^{e}%
\right) ,  \label{26}
\end{equation}%
where we can see that $(i)$ if one identifies the field $e^{a}$ with the
vielbein, the system consists of the Einstein-Hilbert action plus
nonminimally coupled matter fields given by $h^{a}$ and $k^{ab};$ $(ii)$ it
is possible to recover the odd-dimensional Einstein gravity theory from a
Chern-Simons gravity theory in the limit where the coupling constant $l$
equals to zero while keeping the effective Newton's constant fixed.

It is the purpose of this article to find a spherically symmetric solution for the $EChS$ field equations, which are obtained from the so called Einstein-Chern-Simons action (\ref{26}) studied in Refs, \cite{salg1}, \cite{salga1}. It is shown that the standard five dimensional solution of the Einstein-Cartan field equations can be obtained, in a certain limit, from the spherically symmetric solution of $EChS$ field equations. The conditions under which these equations admits black hole type solutions are found and the maximal extension and conformal compactification are also studied.  

This paper is organized as follows: In section 2 we find a spherically
symmetric solution for the Einstein-Chern-Simons field equations and then it
is shown that the standard five dimensional solution of the Einstein-Cartan
field equations can be obtained, in a certain limit, from the spherically
symmetric solution of $EChS$ field equations. In section 3 we find the
conditions under which the field equations admits black hole type solutions
and we studied the maximal extension and conformal compactification of such
solutions.  A brief comment and three appendices conclude this work.

\section{Einstein-Chern-Simons field equations for a spherically symmetric
me\-tric}

In this section we consider the field equations for \ the lagrangian $%
L=L_{g}+L_{M}$ , where $L_{g}$ is the Chern-Simons gravity lagrangian $L_{%
\mathrm{ChS}}^{(5)}$\ and $\ L_{M}$ \ is the corresponding matter lagrangian.

In the presence of matter described by the langragian $%
L_{M}=L_{M}(e^{a},h^{a},\omega ^{ab}),$ we have that the field equations
obtained from the action (\ref{26}) are given by \cite{salga1}: 
\begin{align}
\varepsilon _{abcde}R^{cd}T^{e}& =\ 0,  \notag  \label{8-2} \\
\alpha _{3}l^{2}\varepsilon _{abcde}R^{bc}R^{de}& =-\frac{\delta L_{M}}{%
\delta h^{a}},  \notag \\
\varepsilon _{abcde}\left( 2\alpha _{3}R^{bc}e^{d}e^{e}+\alpha
_{1}l^{2}R^{bc}R^{de}+2\alpha _{3}l^{2}D_{\omega }k^{bc}R^{de}\right) & =-%
\frac{\delta L_{M}}{\delta e^{a}},  \notag \\
2\varepsilon _{abcde}\left( \alpha _{1}l^{2}R^{cd}T^{\text{{}}e}+\alpha
_{3}l^{2}D_{\omega }k^{cd}T^{e}+\alpha _{3}e^{c}e^{d}T^{e}+\alpha
_{3}l^{2}R^{cd}D_{\omega }h^{e}+\alpha _{3}l^{2}R^{cd}k_{\text{ }%
f}^{e}e^{f}\!\right) & =-\frac{\delta L_{M}}{\delta \omega ^{ab}}.
\end{align}

If $T^{a}=0$ and $k^{ab}=0,$ the equation (\ref{8-2}) can be written in the
form

\begin{align}
de^{a}+\omega _{\text{ }b}^{a}e^{b}& =0,  \notag \\
\varepsilon _{abcde}R^{cd}D_{\omega }h^{e}& =0,  \notag \\
\alpha _{3}l^{2}Y_{a}& =-\star \left( \frac{\delta L_{M}}{\delta h^{a}}%
\right) ,  \notag \\
\alpha _{1}l^{2}Y_{a}+2\alpha _{3}X_{a}& =\kappa T_{ab}e^{b},  \label{30-1}
\end{align}%
where%
\begin{equation}\label{30-2}
X_{a}=\star \left( \varepsilon _{abcde}R^{bc}e^{d}e^{e}\right) ,\quad
Y_{a}=\star \left( \varepsilon _{abcde}R^{bc}R^{de}\right) ,\quad
T_{ab}=-\star \left( \frac{\delta L_{M}}{\delta e^{a}}\right) 
\end{equation}%
and where \textquotedblleft $\star $\textquotedblright\ is the Hodge star
operator.

$T_{ab}$ is the energy-momentum tensor of matter fields and $\kappa $ is the
coupling constant. In the equations (\ref{30-1}) are present the fields $%
e^{a}$, $\omega ^{ab}$ (through $R^{ab}$) and $h^{a}$. If we wish to find a
spherically-and static-symmetric solution, then we must demand that the
three fields satisfy this conditions. Since a static space-time is one
which posseses a timelike Killing vector orthogonal to the spacelike
hypersurfaces. These conditions are satisfied by the metric (\ref{31}).

\subsection{\textbf{Spherically symmetric metric in five dimensions}}

\bigskip We consider first the fields $e^{a}$ and $\omega ^{ab}$ (through $%
R^{ab}$). In five dimensions the static and spherically symmetric metric is
given by 
\begin{equation}
ds^{2}=-e^{2f(r)}dt^{2}+e^{2g(r)}dr^{2}+r^{2}d\Omega _{3}^{2}=\eta
_{ab}e^{a}e^{b}  \label{31}
\end{equation}%
where $d\Omega _{3}^{2}=d\theta _{1}^{2}+\sin ^{2}\theta _{1}d\theta
_{2}^{2}+\sin ^{2}\theta _{1}\sin ^{2}\theta _{2}d\theta _{3}^{2}$ and $\eta
_{ab}=\mathrm{diag}(-1,+1,+1,+1,+1)$.

Introducing an orthonormal basis, we have
\begin{align}
e^{T}=e^{f(r)}dt,\quad e^{R}=e^{g(r)}dr,\quad e^{1}=rd\theta_{1},\quad e^{2}=r\sin\theta_{1}d\theta_{2},\quad e^{3}=r\sin\theta_{1}\sin\theta_{2}d\theta_{3}. \label{32}%
\end{align}
Taking the exterior derivatives, we get:
\begin{align}
de^{T}&=-f^{\prime}e^{-g}e^{T}e^{R},\qquad de^{R}=0,\qquad de^{1}=\frac{e^{-g}}{r}e^{R}e^{1},\nonumber\\
de^{2}&=\frac{1}{r\tan\theta_{1}}e^{1}e^{2}+\frac{e^{-g}}{r}e^{R}e^{2},\qquad
de^{3}=\frac{1}{r\tan\theta_{1}}e^{1}e^{3}+\frac{1}{r\sin\theta_{1}\tan\theta_{2}}e^{2}e^{3}+\frac{e^{-g}}{r}e^{R}e^{3},\label{treintay tres}
\end{align}
where a prime \textquotedblleft $\phantom{i}'$ \textquotedblright\ denotes derivative with respect to $r$. The next step is to use Cartan's first structural equation
\[
T^{a}=de^{a}+\omega_{\phantom{2}b}^{a}e^{b}=0
\]
and the antisymmetry of the connection forms ($\omega^{ab}=-\omega^{ba}$) to
find the non-zero connection forms. The calculations give:
\begin{align}
\omega_{\text{ }TR}&=-f^{\prime}e^{-g}e^{T},\qquad\omega_{Ri}=-\frac{e^{-g}}{r}e^{i},\qquad\omega_{12}=-\frac{1}{r\tan\theta_{1}}e^{2},\nonumber\\
\omega_{13}&=-\frac{1}{r\tan\theta_{1}}e^{3},\qquad
\omega_{23}=-\frac{1}{r\sin\theta_{1}\tan\theta_{2}}e^{3};\qquad i=1,2,3.\label{treintay4}
\end{align}

From Cartan's second structural equation
\[
R_{\phantom{2}b}^{a}=d\omega_{\phantom{2}b}^{a}+\omega_{\phantom{2}c}^{a}\omega_{\phantom{2}b}^{c},
\]
we can calculate the curvature matrix. The non-zero components are
\begin{align}
R^{TR}&=e^{-g}\left(  f^{\prime}g^{\prime}-f^{\prime\prime}-\left(  f^{\prime
}\right)  ^{2}\right)  e^{T}e^{R},\qquad R^{Ti}=-\frac{f^{\prime}e^{-2g}}{r}e^{T}e^{i}\nonumber\\
R^{Ri}&=\frac{g^{\prime}e^{-2g}}{r}e^{R}e^{i},\qquad R^{ij}=\frac{1-e^{-2g}}{r^{2}}e^{i}e^{j};\qquad i,j=1,2,3 .\label{33}
\end{align}

Introducing (\ref{32}), (\ref{33}) into (\ref{30-2}) we find
\begin{align}
X_{T}   =&\ 12\frac{e^{-2g}}{r^{2}}\left(  g^{\prime}r+e^{2g}-1\right)
e^{T},\nonumber\\
X_{R}   =&\ 12\frac{e^{-2g}}{r^{2}}\left(  f^{\prime}r-e^{2g}+1\right)
e^{R},\nonumber\\
X_{i}    =&\ 4\frac{e^{-2g}}{r^{2}}\Bigl(  -f^{\prime}g^{\prime}r^{2}%
+f^{\prime\prime}r^{2}+\left(  f^{\prime}\right)  ^{2}r^{2}+2f^{\prime}r-2g^{\prime}r-e^{2g}+1\Bigr)  e^{i},\label{34'}
\end{align}
\begin{align}
Y_{T}  =&\ 24\frac{e^{-2g}}{r^{3}}g^{\prime}\left(  1-e^{-2g}\right)
e^{T},\nonumber\\
Y_{R}  =&\ 24\frac{e^{-2g}}{r^{3}}f^{\prime}\left(  1-e^{-2g}\right)
e^{R},\nonumber\\
Y_{i}  =&\ 8\frac{e^{-2g}}{r^{2}}\Bigl(  f^{\prime\prime}+\left(  f^{\prime
}\right)  ^{2}-f^{\prime}g^{\prime}-e^{-2g}f^{\prime\prime}-e^{-2g}\left(
f^{\prime}\right)  ^{2}+3e^{-2g}f^{\prime}g^{\prime}\Bigr)  e^{i}.\label{35'}
\end{align}

Introducing (\ref{32}), (\ref{34'}), (\ref{35'}) into the third equation
(\ref{30-1}) and considering the energy-momentum tensor as the energy-momentum
tensor of a perfect fluid at rest, i.e., $T_{TT}=\rho(r)$ and $T_{RR}=T_{ii}=P(r),$
where $\rho(r)$ and $P(r)$ are the energy density and pressure (for the
perfect fluid), \ we find
\begin{align}
\alpha_{1}l^{2}\frac{e^{-2g}}{r^{3}}g^{\prime}\left(  1-e^{-2g}\right)
+\alpha_{3}\frac{e^{-2g}}{r^{2}}\left(  g^{\prime}r+e^{2g}-1\right)
&=\frac{\kappa}{24}\rho\label{36'}\\
\alpha_{1}l^{2}\frac{e^{-2g}}{r^{3}}f^{\prime}\left(  1-e^{-2g}\right)
+\alpha_{3}\frac{e^{-2g}}{r^{2}}\left(  f^{\prime}r-e^{2g}+1\right)
&=\frac{\kappa}{24}P \label{37'}\\
\alpha_{1}l^{2}\frac{e^{-2g}}{r^{2}}\left(  f^{\prime\prime}+\left(
f^{\prime}\right)  ^{2}-f^{\prime}g^{\prime}-e^{-2g}f^{\prime\prime}%
-e^{-2g}\left(  f^{\prime}\right)  ^{2}+3e^{-2g}f^{\prime}g^{\prime}\right)&\nonumber\\
+\alpha_{3}\frac{e^{-2g}}{r^{2}}\left(  -f^{\prime}g^{\prime}r^{2}%
+f^{\prime\prime}r^{2}+\left(  f^{\prime}\right)  ^{2}r^{2}+2f^{\prime
}r-2g^{\prime}r-e^{2g}+1\right)  &=\frac{\kappa}{8}P \label{38'}%
\end{align}

Now consider the equation (\ref{36'}). After multiplying by
$4r^{3}$ we find
\begin{equation}
\Bigl\{  \left(  1-e^{-2g}\right)  \Bigl(  \alpha_{1}l^{2}\left(
1-e^{-2g}\right)  +2\alpha_{3}r^{2}\Bigr)\Bigr\}  ^{\prime}=\frac{\kappa
}{6}\rho r^{3}. \label{39'}%
\end{equation}

Integrating we have
\begin{equation}
\left(  1-e^{-2g}\right)\Bigl(  \alpha_{1}l^{2}\left(
1-e^{-2g}\right)  +2\alpha_{3}r^{2}\Bigr)  =\frac{\kappa}{12\pi^{2}}\Bigl(\mathcal{M}(r)-\mathcal{M}_{0}\Bigr),
\label{40'}%
\end{equation}
where $\mathcal{M}_{0}$ is an integration constant and $\mathcal{M}(r)$ is the Newtonian mass,
which is defined as%
\begin{equation}
\mathcal{M}(r)=2\pi^{2}\int_{0}^{r}\rho(\bar{r})\bar{r}^{3}d\bar{r}. \label{41'}%
\end{equation}

From equation (\ref{40'}) we can see that
\begin{equation}
e^{-2g}=1+\alpha\frac{r^{2}}{l^{2}}\pm\sqrt{\alpha^{2}\frac{r^{4}}{l^{4}%
}+\frac{K}{12\pi^{2}l^{2}}\Bigl(  \mathcal{M}(r)-\mathcal{M}_{0}\Bigr)  }, \label{42'}%
\end{equation}
where $\alpha=\alpha_{3}/\alpha_{1},$ \ $K=\kappa/\alpha_{1}.$

In order to make contact with the solutions of the Einstein-Cartan theory,
consider the limit $l\rightarrow0$:%
\begin{equation}
\lim_{l\rightarrow0}e^{-2g}=\lim_{l\rightarrow0}\left(  1+\alpha
\frac{r^{2}}{l^{2}}\pm\sqrt{\alpha^{2}\frac{r^{4}}{l^{4}}+\frac{K}{12\pi
^{2}l^{2}}\Bigl(  \mathcal{M}(r)-\mathcal{M}_{0} \Bigr)}\ \right).  \label{43'}%
\end{equation}

%Since
%\begin{equation}
%\pm\sqrt{\alpha^{2}\frac{r^{4}}{l^{4}}+\frac{K}{12\pi^{2}l^{2}}\left(
%m(r)-m_{0}\right)  }=\left\vert \alpha\right\vert \frac{r^{2}}{l^{2}}\left(
%\pm\sqrt{1+\frac{Kl^{2}}{12\pi^{2}\alpha^{2}r^{4}}\left(  m(r)-m_{0}\right)
%}\right)  \label{44'}%
%\end{equation}
If we consider the case of small $l^{2}$ limit, we can expand the root to
first order in $l^{2}$. \ In fact,%
\begin{align}
e^{-2g}  &  \approx1+\frac{r^{2}}{l^{2}}\left\{  \alpha\pm\left\vert
\alpha\right\vert \left(  1+\frac{Kl^{2}}{12\pi^{2}l^{2}\alpha^{2}r^{4}%
}\Bigl(  \mathcal{M}(r)-\mathcal{M}_{0}\Bigr)  +O(l^{4})\right)  \right\} \nonumber\\
  &  \approx1+\frac{r^{2}}{l^{2}}\left(  \alpha\pm\left\vert
\alpha\right\vert \right)  \pm\frac{K}{24\pi^{2}\left\vert \alpha\right\vert
r^{2}}\Bigl( \mathcal{M}(r)-\mathcal{M}_{0}\Bigr)  +O(l^{4}) .\label{45'}%
\end{align}

From (\ref{45'}) we can see that for this expression to be finite when
$l\rightarrow0,$ is necessary that $\left(  \alpha\pm\left\vert
\alpha\right\vert \right)  =0.$

Since $\alpha=\alpha_{3}/\alpha_{1}$ we can distinguish two cases:
\begin{enumerate}
\item[$(a)$] If $\alpha_{3}>0$ and $\alpha_{1}>0$ or if $\alpha_{3}<0$ and
$\alpha_{1}<0$ we have
\begin{align}
e^{-2g}  &  =1+\alpha\frac{r^{2}}{l^{2}}-\sqrt{\alpha^{2}\frac{r^{4}}{l^{4}%
}+\frac{K}{12\pi^{2}l^{2}}\Bigl(  \mathcal{M}(r)-\mathcal{M}_{0}\Bigr)  }\nonumber\\
&  \approx 1-\frac{K}{24\pi^{2}\left\vert \alpha\right\vert r^{2}}\Bigl(\mathcal{M}(r)-\mathcal{M}_{0}\Bigr) \nonumber\\
&  \approx 1-\frac{\kappa}{24\pi^{2}\alpha_{3}r^{2}}\Bigl( \mathcal{M}(r)-\mathcal{M}_{0}\Bigr).\label{46'}
\end{align}

\item[$(b)$] If $\alpha_{3}>0$ and $\alpha_{1}<0$ or if $\alpha_{3}<0$ and
$\alpha_{1}>0$ we have
\begin{align}
e^{-2g}  &  =1+\alpha\frac{r^{2}}{l^{2}}+\sqrt{\alpha^{2}\frac{r^{4}}{l^{4}%
}+\frac{K}{12\pi^{2}l^{2}}\Bigl(\mathcal{M}(r)-\mathcal{M}_{0}\Bigr)  }\nonumber\\
&  \approx 1+\frac{K}{24\pi^{2}\left\vert \alpha\right\vert r^{2}}\Bigl(\mathcal{M}(r)-\mathcal{M}_{0}\Bigr) \nonumber\\
&  \approx 1-\frac{\kappa}{24\pi^{2}\alpha_{3}r^{2}}\Bigl(\mathcal{M}(r)-\mathcal{M}_{0}\Bigr).
\label{47'}
\end{align}
\end{enumerate}

This means that whatever the choice of the sign of the constant $\alpha_{1}$
and $\alpha_{3}$ we obtain%
\begin{equation}
\lim_{l\rightarrow0}e^{-2g}=1-\frac{\kappa}{24\pi^{2}\alpha_{3}r^{2}%
}\Bigl(\mathcal{M}(r)-\mathcal{M}_{0}\Bigr).  \label{48'}%
\end{equation}
From (\ref{48'}) we can see that if $\kappa/2\alpha_{3}=\kappa_{E}$ and
$\mathcal{M}_{0}=0$ we recover the usual 5-dimensional expresion for $e^{-2g}$ (see
\ref{A19}).

\subsection{The Exterior Solution}

The third equation (\ref{30-1}) can be rewritten in the form%
\begin{equation}
\star\left(  \varepsilon_{abcde}R^{bc}e^{d}e^{e}\right)  +\frac{1}{2\alpha
}l^{2}\star\left(  \varepsilon_{abcde}R^{bc}R^{de}\text{\ }\right)
=\text{\ }\kappa_{E}T_{ab}e^{b}, \label{58'}%
\end{equation}
where $\alpha=\alpha_{3}/\alpha_{1}$ and \ $\kappa_{E}=\kappa/2\alpha_{3}$.

Rescaling the parameter $l$ in the form $l\longrightarrow l^{\prime}=l/\sqrt{\left\vert
\alpha\right\vert }$ we have

\begin{equation}
\star\left(  \varepsilon_{abcde}R^{bc}e^{d}e^{e}\right)  +\textrm{sgn}(\alpha
)\frac{l^{2}}{2}\star\left(  \varepsilon_{abcde}R^{bc}R^{de}\text{\ }\right)
=\text{\ }\kappa_{E}T_{ab}e^{b}. \label{59'}%
\end{equation}%

If $\rho(r)=P(r)=0$ and $\delta L_{M}/\delta h^{a}\neq0,$ the field equations
are given by:
\begin{align}
\frac{e^{-2g}}{r^{3}}g^{\prime}\left(  1-e^{-2g}\right)  +\frac{\textrm{sgn}(\alpha)}{l^{2}%
}\frac{e^{-2g}}{r^{2}}\left(  g^{\prime}r+e^{2g}-1\right)  &=0,\label{61}\\
\frac{e^{-2g}}{r^{3}}f^{\prime}\left(  1-e^{-2g}\right)  +\frac{\textrm{sgn}(\alpha)}{l^{2}%
}\frac{e^{-2g}}{r^{2}}\left(  f^{\prime}r-e^{2g}+1\right)  &=0,\label{62}\\
\frac{e^{-2g}}{r^{2}}\left(  f^{\prime\prime}+\left(  f^{\prime}\right)
^{2}-f^{\prime}g^{\prime}-e^{-2g}f^{\prime\prime}-e^{-2g}\left(  f^{\prime
}\right)  ^{2}+3e^{-2g}f^{\prime}g^{\prime}\right)\qquad\quad&\nonumber\\
+\frac{\textrm{sgn}(\alpha)}{l^{2}}\frac{e^{-2g}}{r^{2}}\left(  -f^{\prime}g^{\prime}%
r^{2}+f^{\prime\prime}r^{2}+\left(  f^{\prime}\right)  ^{2}r^{2}+2f^{\prime
}r-2g^{\prime}r-e^{2g}+1\right)  &=0.\label{63}
\end{align}

Following the usual procedure, we find that the equation (\ref{61}) has the
following solution:
\begin{equation}
e^{-2g}=1+\textrm{sgn}(\alpha)\frac{r^{2}}{l^{2}}-\textrm{sgn}(\alpha)\sqrt{\frac{r^{4}}{l^{4}%
}+\textrm{sgn}(\alpha)\frac{\kappa_{E}}{6\pi^{2}l^{2}}M},\label{64}%
\end{equation}
where $M$ is a constant of integration. From (\ref{64}) is straightforward
to see that in the limit $l\rightarrow0$ we obtain the solution
(\ref{A24}) to Einstein's gravity.

Adding equations (\ref{61}) and (\ref{62}) we find
\begin{equation}
e^{2f}=e^{-2g}.\label{65}%
\end{equation}
This solution satisfies the equation (\ref{63}).

From (\ref{64}) and (\ref{65}) we can see that the line element for the outer
region is given by%
\begin{equation}
ds^{2}=-F(r)dt^{2}+\frac{dr^{2}}{F(r)}+r^{2}d\Omega_{3}^{2}, \label{66}%
\end{equation}
where%
\begin{equation}
F(r)=1+\textrm{sgn}(\alpha)\frac{r^{2}}{l^{2}}-\textrm{sgn}(\alpha)\sqrt{\frac{r^{4}}{l^{4}%
}+\textrm{sgn}(\alpha)\frac{\kappa_{E}}{6\pi^{2}l^{2}}M}.\label{67}%
\end{equation}

\section{Black-Hole solution of Einstein-Chern-Simons field equations}

Let us consider now the conditions under which the equation (\ref{59'}) admits
black hole type solutions.

\subsection{Case $\alpha>0$: Black Holes}

In this case the exterior solution is given by (\ref{66}) with
\begin{equation}
F(r)=1+\frac{r^{2}}{l^{2}}-\sqrt{\frac{r^{4}}{l^{4}}+\frac{\kappa_{E}}%
{6\pi^{2}l^{2}}M}. \label{61'}%
\end{equation}

This solution shows an anomalous behaviour at
\[
F(r_{0})=1+\frac{r_{0}^{2}}{l^{2}}-\sqrt{\frac{r_{0}^{4}}{l^{4}}+\frac
{\kappa_{E}}{6\pi^{2}l^{2}}M}=0,
\]
i.e., at
\begin{equation}
r_{0}=\sqrt{\frac{\kappa_{E}}{12\pi^{2}}M-\frac{l^{2}}{2}} \label{62'}%
\end{equation}
so that
\begin{equation}
F(r)=1+\frac{r^{2}}{l^{2}}-\sqrt{\frac{r^4 +2r_{0}^{2}l^{2}+l^{4}}{l^4}}%
.\label{62''}%
\end{equation}

From the equations (\ref{66}) and (\ref{62'}) we can see that if
\begin{equation}
\frac{\kappa_{E}}{6\pi^{2}}M>l^{2},\label{63'}%
\end{equation}
then the metric (\ref{66}) shows an anomalous behaviour at $r=r_{0}.$ A first
elementary anomaly is that we have at $r=r_{0}$
\begin{equation}
g_{00}=g^{11}=0;\qquad g^{00}=g_{11}=\infty.\label{64'}%
\end{equation}

A more serious anomaly is the following. One can verify that the
parametric lines of the coordinate $r$, i.e. the lines on which the
coordinates $t,\theta_{1},\theta_{2},\theta_{3}$ have constant
values, are geodesics. But these geodesics are space-like for $r>r_{0}$ and
time-like for $r<r_{0}.$ \ The tangent vector of a geodesic undergoes
parallel transport along the geodesic and consequently it cannot change from a
time like to a space-like vector. It follows that the two regions \ $r>r_{0}$
and $r<r_{0}$ do not joint smoothly on the surface $r=r_{0}%
$.

This can be see in a more striking manner if we consider the radial
null directions, on which $d\theta_{1}=d\theta_{2}=d\theta_{3}=0$. We have
then
\begin{equation}
ds^{2}=-F(r)dt^{2}+\frac{dr^{2}}{F(r)}=0.\label{65'}%
\end{equation}

Consequently the radial null directions satisfies the relations 
\begin{equation}
\frac{dr}{dt}=\pm F(r). \label{66'}%
\end{equation}

\begin{figure}[t]
\includegraphics[width=0.6\columnwidth]{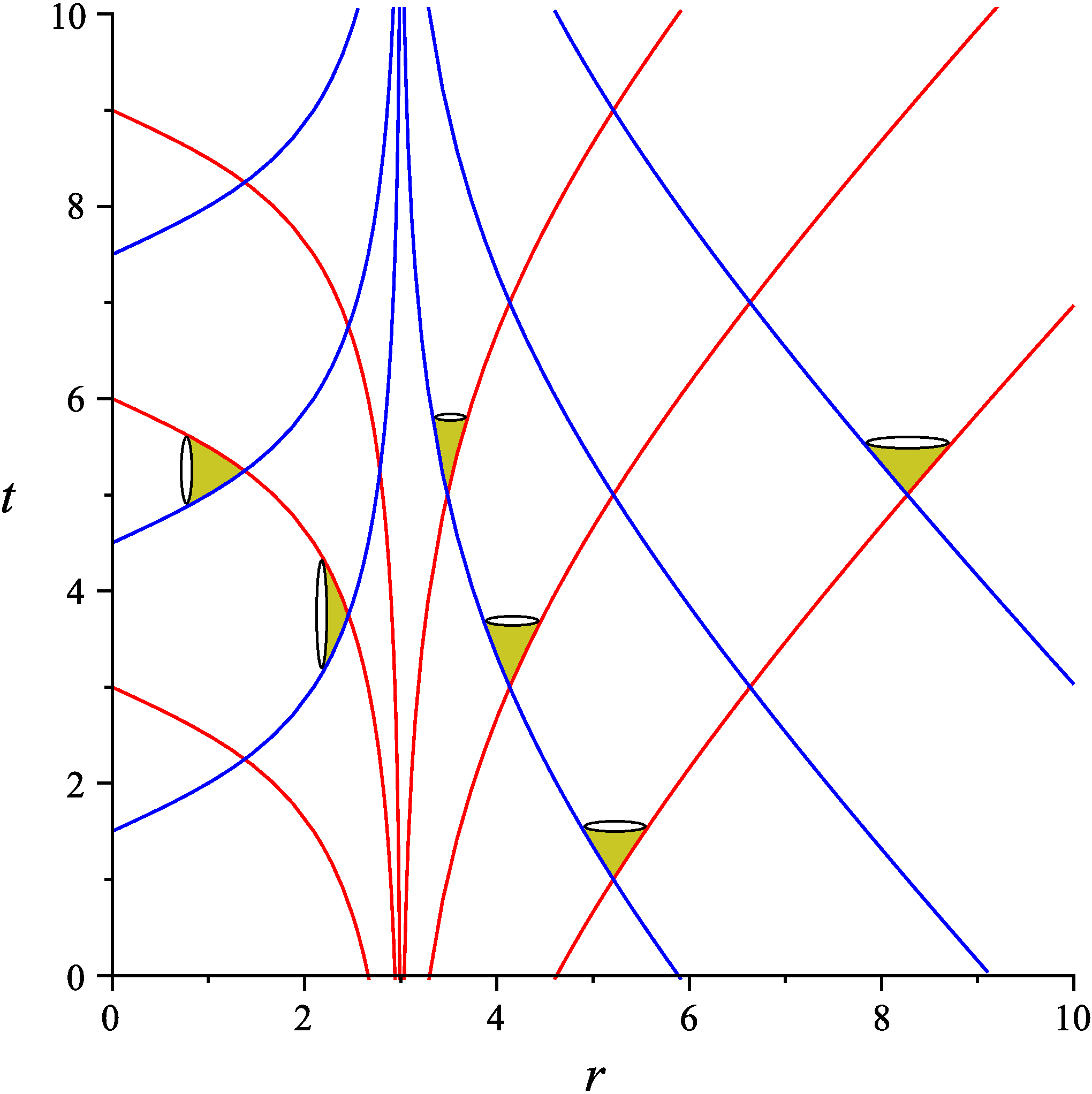}
\centering
\caption{\small{Space-time diagram in Schwarzschild-like coordinates for $l^2=2$ and $\kappa_{EC}(6\pi^2)^{-1}M=20$, so that $r_0=3$. Some future light cone has been drawn.\label{fig4.3}}}
\end{figure}

If we take into account the fact that the time-like directions are
contained in the light-cone, we find that in the region $r>r_{0}$ the
light cones have, in the plane $\left(  r,t\right)  $, the
orientation shown on the figure \ref{fig4.3}.

The opening of the light cone, which is nearly equal to $\pi
/4$ for $r\gg r_{0}$, decreases with $r$ and tends to
zero when $r\rightarrow r_{0}$. On the contrary, in the region
$r<r_{0}$ the parametric lines of the coordinate $t$ are
space-like and consequently the light cones are oriented as shown on the
left-hand side of figure \ref{fig4.3}, the opening of the cone increasing from the
value zero at $r=0$ to $\pi/2$ at $r=r_{0}$.
 Comparing the two diferent forms of the light cones on figure (\ref{fig4.3}), we see
that the regions on either side of the surface $r=r_{0}$ do not join
smoothly on this surface.

\subsection{Eddington-Finkelstein and Kruskal-Szekeres coordinates}

Let us define a radial coordinate 

\begin{equation}
r^{\ast}=\int\frac{dr}{F(r)}, \label{67'}%
\end{equation}
we obtain (see appendix \ref{int01})
\begin{align}
r^{*} =\ &\frac{r}{2}+\frac{r_0^2+l^2}{4r_0}\Biggl(\ln\left(\frac{(r-r_0)^2}{r_0(r+r_0)}\right)+ \mathrm{Z}_{\alpha>0}(r)\Biggr)\nonumber\\
&-\frac{ir_0^2}{2}\sqrt{\frac{i}{\sqrt{2r_0^2l^2+l^4}}}\ \mathrm{F}\left(\sqrt{\frac{i}{\sqrt{2r_0^2l^2+l^4}}}\ r,i\right)\nonumber\\
& + \frac{1}{2}\sqrt{i\sqrt{2r_0^2l^2+l^4}}\Biggl\{ \mathrm{F}\left(\sqrt{\frac{i}{\sqrt{2r_0^2l^2+l^4}}}\ r,i\right)-\mathrm{E}\left(\sqrt{\frac{i}{\sqrt{2r_0^2l^2+l^4}}}\ r,i\right)\Biggr\}. \label{75} 
\end{align}

In these coordinates the equation of the null geodesic (\ref{66'}) takes the form 
\begin{equation}
d(t\pm r^{\ast})=0.
\end{equation}
This means 
\begin{equation}
\frac{dt}{dr}=\pm\frac{dr^{\ast}}{dr},
\end{equation}
so that
\begin{equation}
t=\pm r^{\ast}+C_{\pm}.
\end{equation}

The constant $C_{+}$ ($C_{-}$)  uniquely tells us when a photon was sent away (towards) the horizon. We can therefore, consider $v\equiv t+r^{\ast}$ as a new time coordinate,
which brings the metric on the form 
\begin{equation}
ds^{2}=-F(r)dt^{2}+2dvdr+r^{2}d\Omega_{3}^{2}. \label{68'}%
\end{equation}
We now have a non-singular description of particles falling inwards
towards $r=0$ from spatial infinity ($r=\infty$). These
coordinates are called \emph{ingoing Eddington-Finkelstein-coordinates. }

Likewise, if we had chosen $u\equiv t-r^{\ast}$ as a new time
coordinate we would have gotten the metric
\begin{equation}
ds^{2}=-F(r)dt^{2}-2dudr+r^{2}d\Omega_{3}^{2}. \label{69'}%
\end{equation}
These coordinates have a non-singular description of particles
travelling outwards. 

\begin{figure}[t]
\includegraphics[width=0.6\columnwidth]{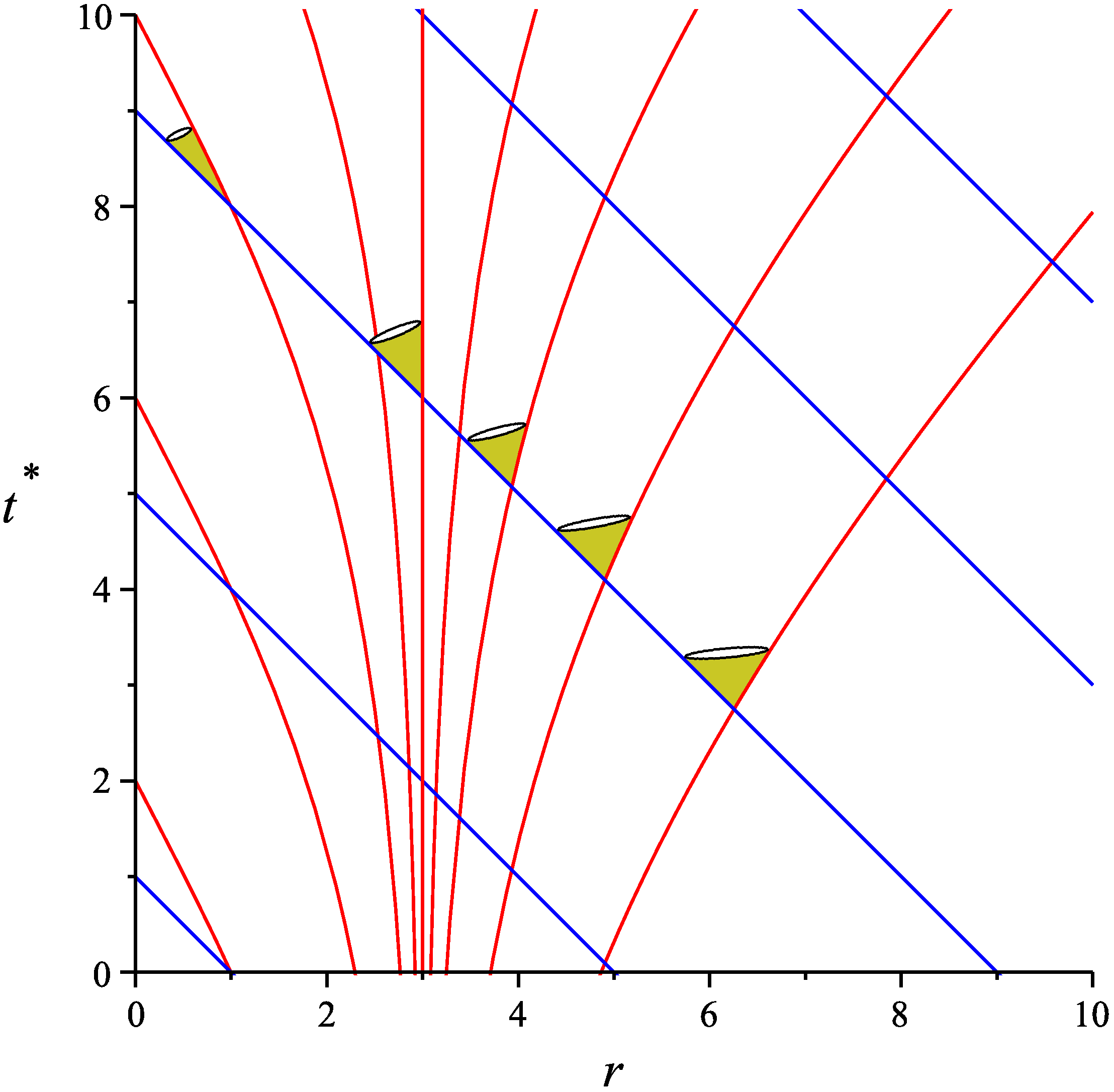}
\centering
\caption{\small{Space-time diagram in advanced Eddington-Finkelstein coordinates for $l^2=2$ and $r_0=3$. Some future light cone has been drawn.\label{fig4.4}}}
\end{figure}

To understand the causal structure in the vicinity of $r=r_{0}$ is useful to define a new timelike coordinate. In effect, let us define
\begin{equation}
t^{\ast}\equiv v-r,
\end{equation}
so that the ingoing null geodesics are given by
\begin{equation}
t^{\ast}=-r+C_{-}.
\end{equation}
These are the straight parallel lines shown on figure \ref{fig4.4}. The outgoing null geodesics are
\begin{equation}
t^{\ast}=2r^{\ast}-r+C_{+}.
\end{equation}

We now recall that physical particles move on time-like worldlines or
on null-lines, i.e. on lines which lie inside or on the surface of the light
cones. It follows then from figure \ref{fig4.4} first of all
that no particle can cross the surface $r=r_{0}$ outwards. Moreover,
any particle which is at some moment inside the surface $r=r_{0}$
will necessarily move towards the singularity in $r=0$, reaching it in finite
coordinate time as well as proper time.

The fact that no particle can cross the surface $r=r_{0}$
outwards means that any observer situated in the region $r>r_{0}$
cannot receive any information about events occurring inside the surface
$r=r_{0}$. We say that the surface $r=r_{0}$ is an (event)
horizon for all observers in the region $r>r_{0}$.  

\begin{figure}[t]
\includegraphics[width=0.6\columnwidth]{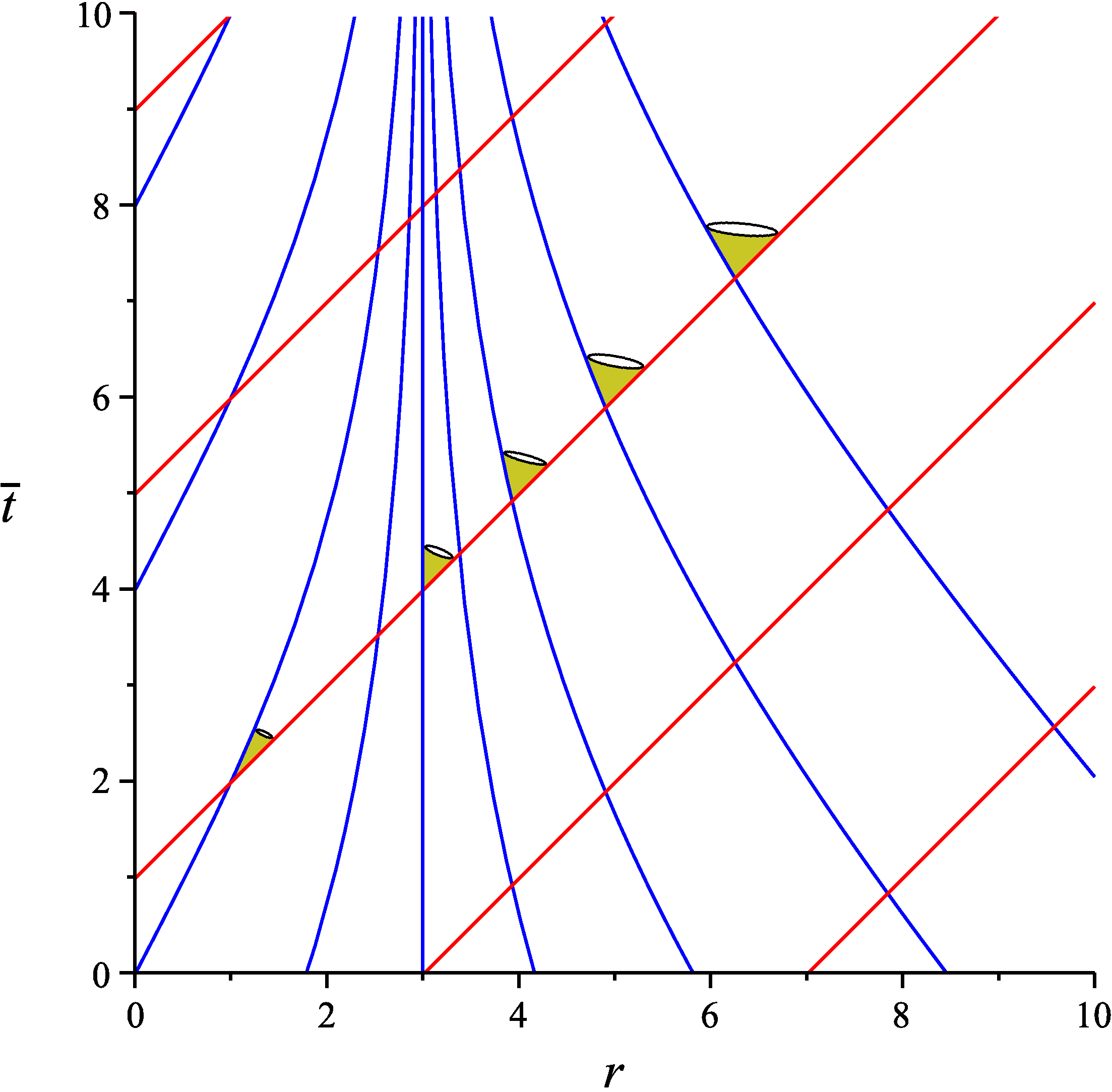}
\centering
\caption{\small{Space-time diagram in retarded Eddington-Finkelstein coordinates. $\bar{t}\equiv u+r$.\label{fig4.5}}}
\end{figure}

From the metric (\ref{69'}) we can see that in this case we shall have
instead of figure \ref{fig4.4} the new figure \ref{fig4.5} resulting from the preceding one by reflexion with respect to the axis $\vec{Or}$.

We see from figure \ref{fig4.5} that now no particles can
cross the horizon inward and that particles situated at some moment in the
region $r<r_{0}$ will necessarily move outwards and reach the horizon
in finite proper time.

The coordinates used in (\ref{68'}) and (\ref{69'}) have, compared
with those used in (\ref{66}), the advantage that they describe the
neighbourhood of the surface $r=r_{0}$ in a satisfactory way.
However, the metrics (\ref{68'}) and (\ref{69'}) has still a certain
deficiency the same type that appears in the Schwarzschild solution of general
relativity. This deficiency is avoided in the Kruskal coordinates, which describe
a geodesically complete space. The new coordinates are defined by choosing the
combination 

\begin{equation}
u =t-r^{\ast},\qquad v =t+r^{\ast};\qquad -\infty<u,v<\infty\label{70'}
\end{equation}
i.e.,
\begin{equation}\label{71'}
t =\frac{1}{2}\left(  u+v\right),\qquad r^{\ast} =\frac{1}{2}\left(  v-u\right),
\end{equation}
so that the metric for $r>r_{0}$ is given by
\begin{equation}
ds^{2}=-F(r)dudv+r^{2}d\Omega_{3}^{2}. \label{72'}%
\end{equation}

This does not quite take care of the problem at the horizon. However,
introducing the Kruskal-Szekeres coordinates 
\begin{equation}
U=-\exp\left(  -\frac{u}{2\beta}\right),\qquad V=\exp\left(
-\frac{v}{2\beta}\right);\qquad r>r_{0},\label{73'}
\end{equation}
where $\beta$ is a parameter which will be determined, we get
the result
\begin{equation}
ds^{2}=-\frac{4\beta F(r)}{\exp\left(  \frac{r^{\ast}}{\beta}\right)
}dudv+r^{2}d\Omega_{3}^{2}. \label{74'}%
\end{equation}

Using the expresion for $r^{\ast}$ given in the equation (\ref{75}), we
obtain
\begin{align}
\exp\left(  \frac{r^{\ast}}{\beta}\right)   &  =\frac{(r-r_0)^{\frac{r_0^2+l^2}{2\beta r_0}}}{\Bigl(r_0(r+r_0^2)\Bigr)^{\frac{r_0^2+l^2}{4\beta r_0}}}\exp\Biggl[\frac{r}{2\beta}+\frac{r_0^2+l^2}{4\beta r_0} \mathrm{Z}_{\alpha>0}(r) \nonumber\\
&  -\frac{ir_0^2}{2\beta}\sqrt{\frac{i}{\sqrt{2r_0^2l^2+l^4}}}\ \mathrm{F}\left(\sqrt{\frac{i}{\sqrt{2r_0^2l^2+l^4}}}\ r,i\right)
\label{75'}\\
&  +\frac{1}{2\beta}\sqrt{i\sqrt{2r_0^2l^2+l^4}}\Biggl\{ \mathrm{F}\left(\sqrt{\frac{i}{\sqrt{2r_0^2l^2+l^4}}}\ r,i\right)-\mathrm{E}\left(\sqrt{\frac{i}{\sqrt{2r_0^2l^2+l^4}}}\ r,i\right)\Biggr\}\Biggr]. \nonumber
\end{align}
Note that the term $\left( r-r_{0}\right)^{\frac{r_{0}^{2}+l^2}{2\beta r_{0}}}$ is responsible for the term $\exp\left(\frac{r^{\ast}}{\beta}\right)$ becomes zero or becomes divergent at
$r=r_{0}$.

Now consider the function $F(r)$ given in (\ref{62''})
\begin{equation}\label{76}
F(r) =1+\frac{r^{2}}{l^{2}}-\sqrt{\frac{r^{4}+2r_0^2l^2+l^4}{l^{4}}}.\nonumber
\end{equation}
Expanding $F(r)$ in power series about $r=r_{0}$, we have
\begin{equation}
F(r)=(r-r_{0})\left(  \frac{2r_{0}}{r_{0}^{2}+l^2}+O(r-r_{0})\right).
\label{77}%
\end{equation}

From (\ref{75'}) and (\ref{77}) we can see if $\beta=\frac{r_{0}^{2}+l^2}{2r_{0}}$ 
then the term $F(r)/\exp\left(  \frac{r^{\ast}}{\beta}\right)  $ is not null or divergent. So
that the line element is given by
\begin{equation}
ds^{2}=-\frac{2\left( r_{0}^{2}+l^2\right)  F(r)}{r_{0}\exp\left(
\frac{2r_{0}r^{\ast}}{r_{0}^{2}+l^2}\right)  }dUdV+r^{2}d\Omega_{3}^{2},
\label{78}%
\end{equation}
where $r>r_{0}$, $U<0$ and $V>0$. 

We can define
\begin{equation}
F_{\alpha>0}(r)=\frac{2\left( r_{0}^{2}+l^2\right)  F(r)}{r_{0}\exp\left(
\frac{2r_{0}r^{\ast}}{r_{0}^{2}+l^2}\right)  },\quad r>0,
\end{equation}
therefore we can let that $U$ and $V$ take any values
\begin{equation}
ds^{2}=-F_{\alpha}(r)dUdV+r^{2}d\Omega_{3}^{2}\quad ,\quad r>0.
\label{81}%
\end{equation}

The curves $U=constant$ and $V=constant$ are null
geodesics. Introducing the Kruskal coordinates, they are given by 
\begin{equation}
T=\frac{1}{2}(U+V),\qquad X=-\frac{1}{2}(U-V), \label{83}%
\end{equation}
which (when $r>r_{0}$) are timelike and spacelike
respectively. 

From (\ref{83}) we can see that
\begin{equation}
UV=T^{2}-X^{2},\qquad\frac{V}{U}=\frac{T+X}{T-X}. \label{84}%
\end{equation}
So that the line element takes the form 
\begin{equation}
ds^{2}=-F_{\alpha}(r)\left(  -dT^{2}+dX^{2}\right)  +r^{2}d\Omega_{3}^{2}.
\label{85}%
\end{equation}

\subsection{Maximal extension and conformal compactification}

Now consider the diagram of the solution (\ref{66}) at coordinates
($X-T$), holding $\theta_{1},\theta_{2}$ and
$\theta_{3}$ fixed, with $X$ the horizontal axis and 
$T$ the vertical axis. 

\paragraph{Consider the curves characterized by $r$ constant:}
\begin{enumerate}
\item[$(i)$] The singularity at $r=0$, correspond to $r^{\ast}=0$, is now two hyperbolas corresponding to the solutions $UV=T^{2}-X^{2}=1$. The manifold is defined only between these two
curves.

\item[$(ii)$] The surfaces of $r=constant>r_{0}$ are
hyperbolas $UV=T^{2}-X^{2}=-b^{2}$, with $b=\exp\left(  r^{\ast
}/2\beta\right)$. The \textquotedblleft asymptotic region\textquotedblright , where $r$ is
very large compared to $r_{0}$ is two regions in the Kruskal
diagram.

\item[$(iii)$] The surfaces of $r=constant$ with
$0<r<r_{0}$ are hyperbolas $UV=T^{2}-X^{2}=b^{2}$, with
$0<b=\exp\left(  r^{\ast}/2\beta\right)  <1$.

\item[$(iv)$] The radius $r=r_{0}$ (the event horizon) is at
$UV=T^{2}-X^{2}=0$, or $T=\pm X$.

\item[$(v)$] If $r\rightarrow\infty$, then $X^{2}
-T^{2}\rightarrow\infty$.

\end{enumerate}

\paragraph{Consider the curves characterized by $t$ constant: }
\begin{enumerate}
\item[$(i)$] If $r>r_{0}$ we have $V/U=-c^{2}$, so that
$T=eX$ with $c=\exp\left(  t/2\beta\right)  $ and
$e=(c^{2}-1)/(c^{2}+1)\in\left[  -1,1\right]$.

\item[$(ii)$] If $0\leq r<r_{0}$ we have $V/U=c^{2}$, so
that $T=e^{\prime}X$ with $e^{\prime}=(c^{2}+1)/(c^{2}-1)\in\left(
-\infty,-1\right)  \cup\left(  1,\infty\right)$.
\end{enumerate}

\begin{figure}[t]
\includegraphics[width=0.6\columnwidth]{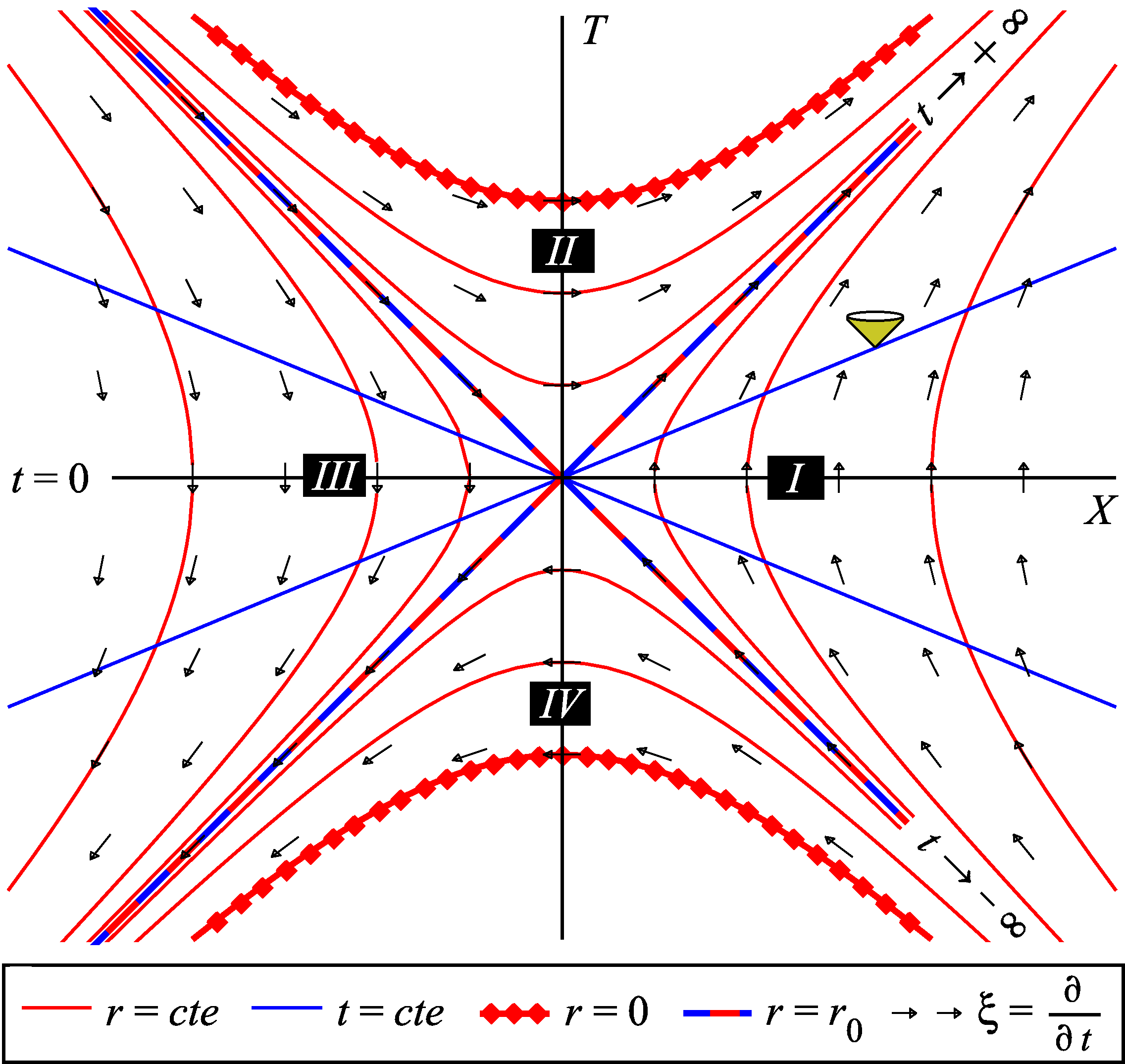}
\centering
\caption{\small{Space-time diagram in Kruskal-Szekeres coordinates that shows its maximal analytic extension.\label{fig4.7}}}
\end{figure}

This means that the surfaces $T=constant$ are
straight lines passing through the origin.

These Kruskal-Szekeres coordinates cover the whole spacetime and show
explicitly that the horizon at $r=r_{0}$ is a mere coordinate
singularity in the Schwarzschild coordinates.

In figure \ref{fig4.7} we have illustrated the Kruskal-Szekeres diagram for
the analytically extended solution (\ref{66}). The original metric covers the region $I$, while region $II$ is the interior of the
black hole. Region $IV$ is the interior of a \textquotedblleft
white hole\textquotedblright\ while region $III$ is just a copy of
region $I$.

Penrose diagrams, or Penrose-Carter diagrams, are a way to represent
the structure of infinity in different spacetimes. A Penrose diagram is a
space-time diagram of a conformally compactified space-time. The idea is to make a coordinate transformation that brings points at infinity in to finite values of the coordinates. Since the angular coordinates (in our case $\theta_{1},\theta_{2},\theta_{3}$) have finite ranges
anyhow, usually we will ignore them and plot one timelike coordinate and one
radial coordinate, or some combination of these.

Of course it is not possible transform an infinite manifold into a
finite region. What is possible is to find a metric that is not the same as
the original one, but related by a conformal transformation. The essential
idea is to start off with a metric $g_{\mu\nu}$, which we call the
physical metric, and introduce another metric $\bar{g}_{\mu\nu},$
called the unphysical metric, which is conformally related to $g_{\mu\nu}%
,$ that is $\bar{g}_{\mu\nu}=\Omega^{2}g_{\mu\nu}$, where
$\Omega$ is the conformal factor. Then, by a suitable choice of
$\Omega^{2},$ it may be possible to \textquotedblleft bring in\textquotedblright\ the points at infinity
to a finite position and hence study the causal structure of infinity.

It is well known that the null geodesics of conformally related
metrics are the same, and that such null geodesics determine the light cones,
which in turn define the causal structure. The essential idea for bringing in
the points at infinity is to use coordinate transformations involving
functions like ${\arctan}(x)$, which, for example, maps the
infinite interval $\left(  -\infty,\infty\right)  $ onto the finite
interval $\left(  -\pi/2,\pi/2\right)  $.

We introduce the null coordinate $q$ and $p$ defined from the Kruskal
coordinates
\begin{equation}
U  =\tan q,\qquad V   =\tan p.\label{86}%
\end{equation}

From (\ref{86}) we can see that if $U\rightarrow\pm\infty$ then $q\rightarrow
\pm\pi/2$ and if $V\rightarrow\pm\infty$ then $p\rightarrow\pm\pi/2.$ \ Now we
introduce a timelike and a spacelike coordinates defined by
\begin{equation}
\tau=p+q,\qquad x=p-q .\label{87}%
\end{equation}

\begin{figure}[t]
\includegraphics[width=0.6\columnwidth]{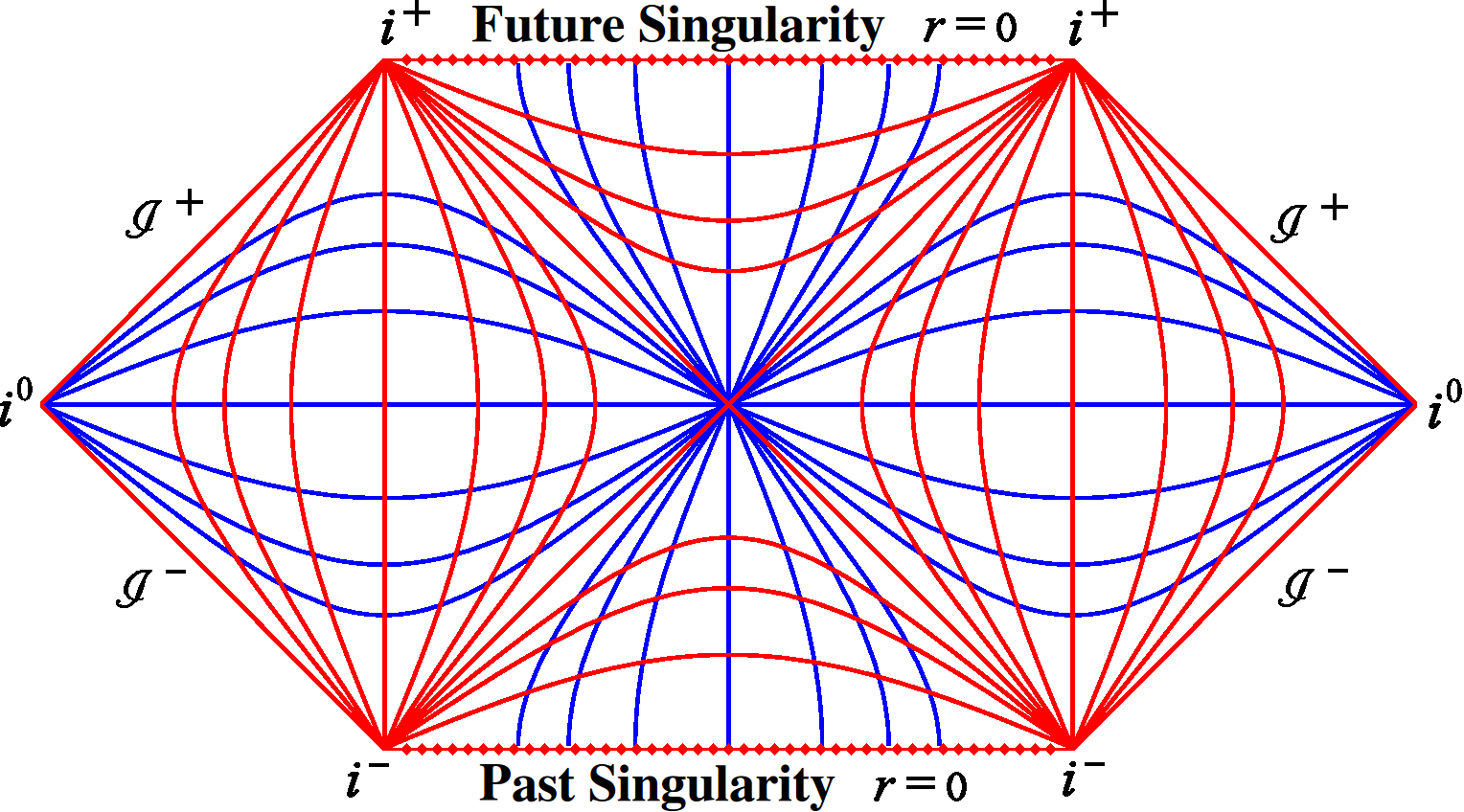}
\centering
\caption{\small{Penrose diagram for $\alpha>0$ with $\frac{\kappa_{E}}{6\pi^{2}}M> l^{2}$.\label{fig4.8}}}
\end{figure}

A space-time diagram at ($x-\tau$) coordinates is shown in figure \ref{fig4.8}.

\paragraph{The curves characterized by $r$ constant are given by}

\begin{enumerate}
\item[$(i)$] If $r=0$ we have
\begin{equation}
UV=\tan p\tan q=\frac{\cos x-\cos\tau}{\cos x+\cos\tau}=1, \label{88}%
\end{equation}
so that $\cos\tau=0.$ This means that $\tau=\pm\pi/2$.

\item[$(ii)$] If $0<r<r_{0}$ we have $UV=b^{2}$ with $0<b<1,$ so that%
\[
\frac{\cos x-\cos\tau}{\cos x+\cos\tau}=b^{2}%
\]
and therefore
\begin{equation}
\cos\tau=\frac{1-b^{2}}{1+b^{2}}\cos x. \label{89}%
\end{equation}

\item[$(iii)$] If $r=r_{0}$ we have $UV=0,$ so that $\cos x-\cos\tau=0$ and
therefore $\tau=\pm x.$

\item[$(iv)$] If $r>r_{0}$ we have $UV=-b^{2}$ with $\ b>0,$ so that%
\[
\frac{\cos x-\cos\tau}{\cos x+\cos\tau}=-b^{2}%
\]
and therefore
\begin{equation}
\cos\tau=\frac{1+b^{2}}{1-b^{2}}\cos x. \label{90}%
\end{equation}

\item[$(v)$] If $r\rightarrow\infty$ we have $UV\rightarrow-\infty$, so that
\begin{equation}
\tau=\pm\pi-x\text\quad\text{or}\quad\tau=\pm\pi+x .\label{91}%
\end{equation}
\end{enumerate}

\textbf{The curves characterized by $t$ constant are given by}

\begin{enumerate}
\item[$(i)$] Since $V/U=c,$ where
\begin{equation}
c=\left\{
\begin{array}
[c]{c}%
\exp\left(  \frac{t}{\beta}\right),\quad 0\leq r\leq r_{0}\\
-\exp\left(  \frac{t}{\beta}\right),\quad r>r_{0}%
\end{array}
\right. , \label{92}%
\end{equation}
we have
\begin{equation}
\frac{V}{U}=\frac{\tan p}{\tan q}=\frac{\sin\tau+\sin x}{\sin\tau-\sin x},
\label{93'}%
\end{equation}
so that%
\begin{equation}
\sin\tau=\frac{c+1}{c-1}\sin x.\label{94'}%
\end{equation}

\item[$(ii)$] If $t=0$ we have $\tau=0$\ or\ $x=0$.

\item[$(iii)$] If $t\rightarrow-\infty$ we have $\tau=-x$.

\item[$(iv)$] If $t\rightarrow\infty$ we have $\tau=x$.
\end{enumerate}

The null geodesics are given by $U=constant$ \ and $V=constant$, so that
$q=constant$ \ and $p=constant.$ This means that
\[
\tau=\pm x+constant.
\]

\subsection{Case $\alpha>0$: Naked singularity}

From the equations (\ref{66}) and (\ref{62'}) we can see that if
\begin{equation}
\frac{\kappa_{E}}{6\pi^{2}}M\leq l^{2}, \label{95}%
\end{equation}
then
\begin{equation}
F(r)=1+\frac{r^{2}}{l^{2}}-\sqrt{\frac{r^{4}}{l^{4}}+\frac{\kappa_{E}}%
{6\pi^{2}l^{2}}M} \label{96}%
\end{equation}
has no real roots.

Defining a radial coordinate\textbf{ }%
\begin{equation}
r^{\ast}=\int\frac{dr}{F(r)} \label{97}%
\end{equation}
\begin{figure}[t]
\includegraphics[width=0.6\columnwidth]{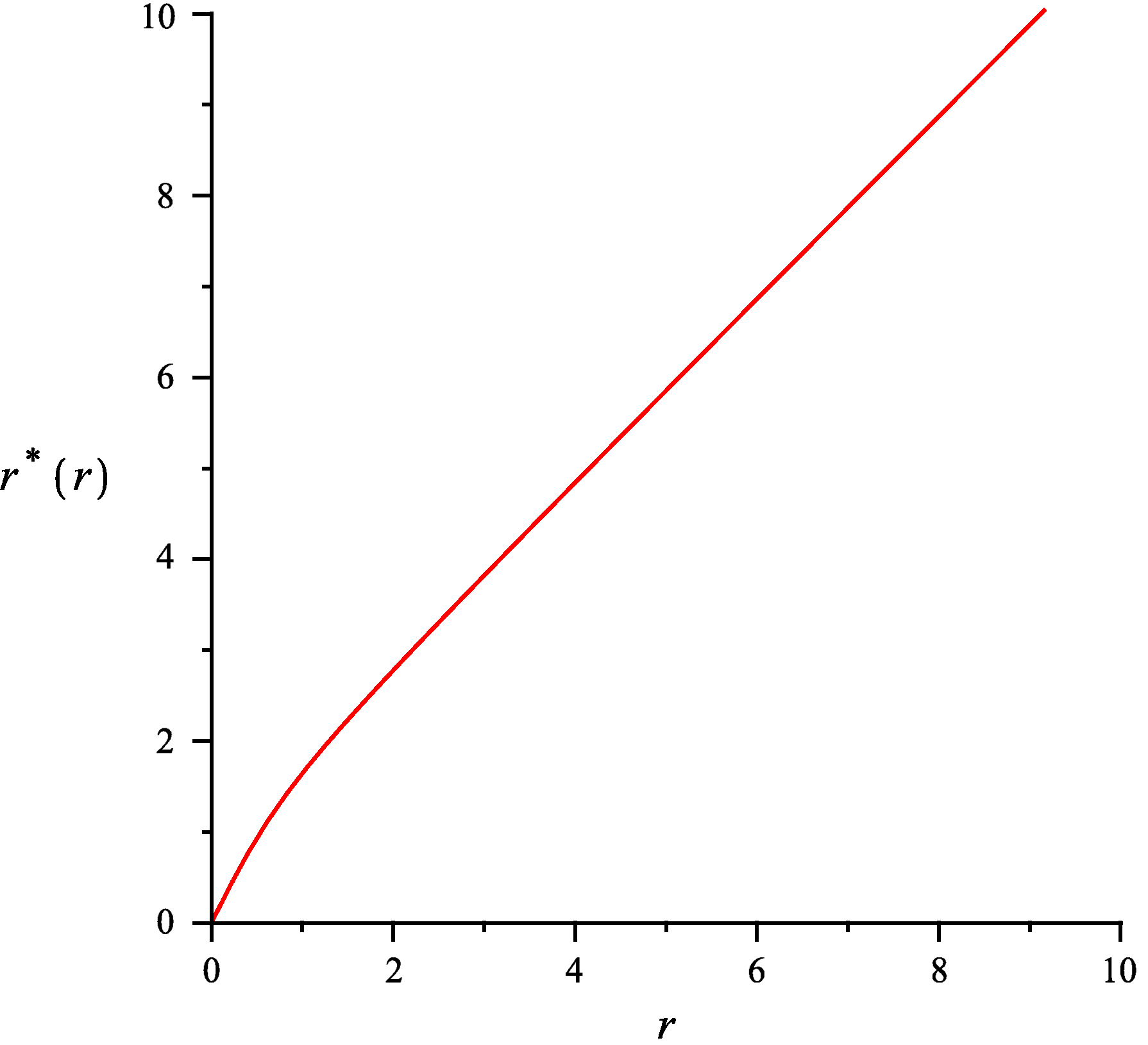}
\centering
\caption{\small{Graph for $r^*(r)$ with $l^2=2$ and $m=1/2$.\label{fig4.9}}}
\end{figure}
Setting the integration constant so that $r^{\ast}(r=0)=0$, we obtain (see appendix \ref{int02})
\begin{align}
r^{\ast}\left(  r\right)   &  =\frac{r}{2}+\frac{m+l^2}{2\sqrt{2(l^2-m)}}%
\arctan\left(  \sqrt{\frac{2}{l^2-m}}\ r\right) \nonumber\\
&  +\frac{1}{2}\sqrt{il\sqrt{m}}\left\{ \mathrm{F}\left(
\sqrt{\frac{i}{l\sqrt{m}}}\ r,i\right)  -\mathrm{E}\left(  \sqrt{\frac
{i}{l\sqrt{m}}}\ r,i\right)  \right\} \label{98}\\
&  +\frac{i(l^2-m)}{4l}\sqrt{\frac{il}{\sqrt{m}}}\mathrm{F}\left(  \sqrt
{\frac{i}{l\sqrt{m}}}\ r,i\right)  -\frac{i\left(m+l^2\right)^{2}}{4l(l^2-m)}\sqrt{\frac
{il}{\sqrt{m}}}\Pi\left(  \sqrt{\frac{i}{l\sqrt{m}}}r,\frac{2il\sqrt{m}}%
{l^2-m},i\right) \nonumber
\end{align}
where $m=\kappa_{E}M/6\pi^{2}$. (see Figure \ref{fig4.9})

The corresponding radial null geodesic, incoming and outgoing, are given
respectively by%
\begin{equation}
t=-r^{\ast}\left(  r\right)  +C,\qquad t=r^{\ast}\left(  r\right)  +C.
\label{99}%
\end{equation}

Since no singularities can be removed, we have no maximal extensions for this solution.

Consider then the corresponding conformal compactification. Let's start by defining the radial null coordinates, incoming and outgoing, as
\begin{equation}
u=t+r^{\ast},\qquad v =t-r^{\ast};\quad-\infty<u,v<\infty. \label{100}%
\end{equation}

\begin{figure}[t]
\includegraphics[width=0.25\columnwidth]{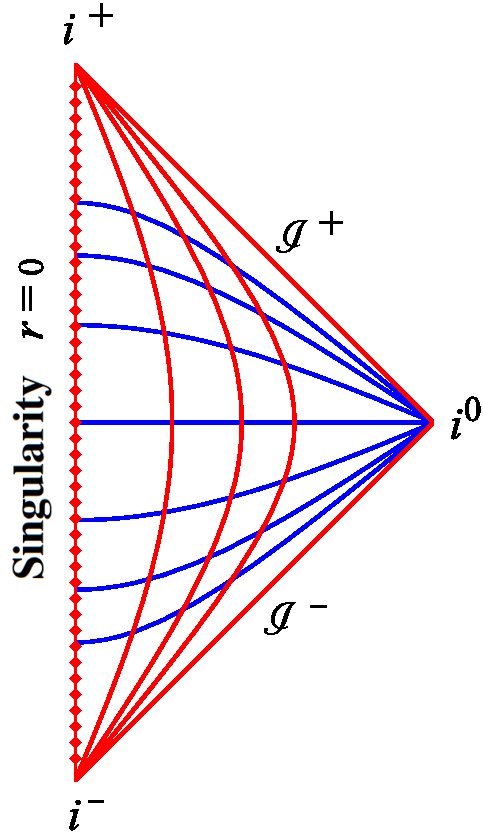}
\centering
\caption{\small{Penrose diagram for the case $\alpha>0$ with $\frac{\kappa_{E}}{6\pi^{2}}M\leq l^{2}$.\label{fig4.10}}}
\end{figure}

The corresponding coordinates, type Kruskal, are defined by
\begin{equation}
U=-\exp\left(  -\frac{u}{2}\right),\qquad V=\exp\left(
\frac{v}{2}\right),\label{101}%
\end{equation}
so that 
\begin{equation}
UV=-\exp\Bigl(  r^{\ast}(r)\Bigr)\quad \text{and}\quad\frac{V}{U}=-\exp(t).
\label{102}%
\end{equation}

Then the line element is given by
\begin{equation}
ds^{2}=-\frac{4F(r)}{\exp\left(  r^{\ast}\right)  }dUdV+r^{2}d\Omega_{3}^{2}.
\label{103}%
\end{equation}

We conducted the compactification, defining the following null coordinate $q$
and $p$
\begin{align}
U  &  =\tan q,\qquad-\frac{\pi}{2}<q\leq0\nonumber;\\
V  &  =\tan p,\qquad 0<p<\frac{\pi}{2} .\label{104}%
\end{align}

Now we introduce the coordinates defined by
\begin{equation}
\tau=p+q\quad\text{and}\quad x=p-q .\label{105}%
\end{equation}

The figure \ref{fig4.10} shows the corresponding Penrose diagram.

\subsection{Case $\alpha<0$: Black Holes}

In this case the exterior solution is given by (\ref{66}) with
\begin{equation}
F(r)=1-\frac{r^{2}}{l^{2}}+\sqrt{\frac{r^{4}}{l^{4}}-\frac{\kappa_{E}}%
{6\pi^{2}l^{2}}M} .\label{106}%
\end{equation}

From (\ref{106}) we can see that there is a minimum value of $r$,
\begin{equation}
r_{m}=\sqrt[4]{\frac{\kappa_{E}Ml^{2}}{6\pi^{2}}} \label{107}%
\end{equation}
for which the function $F(r)$ is well defined. However, it is
straightforward to see that
\begin{align}
R_{\mu\nu\rho\sigma}R^{\mu\nu\rho\sigma}  &  =\frac{4}{l^{4}}\left\{
\frac{4r^{12}}{\left(  r^{4}-r_{m}^{4}\right)  ^{3}}-\frac{12r^{8}}{\left(
r^{4}-r_{m}^{4}\right)  ^{2}}+\frac{4r^{6}}{\left(  r^{4}-r_{m}^{4}\right)
^{\frac{3}{2}}}\right. \nonumber\\
&  \qquad\quad\left. +\frac{15r^{4}}{r^{4}-r_{m}^{4}} -\frac{18r^{2}}{\sqrt{r^{4}-r_{m}^{4}}}-\frac{6\sqrt{r^{4}%
-r_{m}^{4}}}{r^{2}}-\frac{3r_{m}^{4}}{r^{4}}+13\right\},  \label{108'}%
\end{align}
where we see that at $r=r_{m}$, the invariant $R_{\mu\nu\rho\sigma}R^{\mu
\nu\rho\sigma}$ diverges. This means that the 3-sphere defined by $r=r_{m}$ is
a space-time singularity.

From the equations (\ref{66}) and (\ref{62'}) we can see that if
\begin{equation}
\frac{\kappa_{E}}{6\pi^{2}}M>l^{2}, \label{109}%
\end{equation}
then the metric (\ref{66}) shows an anomalous behaviour at
\begin{equation}
r=r_{0}=\sqrt{\frac{\kappa_{E}}{12\pi^{2}}M+\frac{l^{2}}{2}}=\sqrt{\frac
{r_{m}^{4}+l^{4}}{2l^{2}}}. \label{110}%
\end{equation}

Analogously to the case $\alpha>0$, we consider the analysis of causal
structure of spacetime in the vicinity of $r=r_{0}$.

\begin{figure}[t]
\includegraphics[width=0.6\columnwidth]{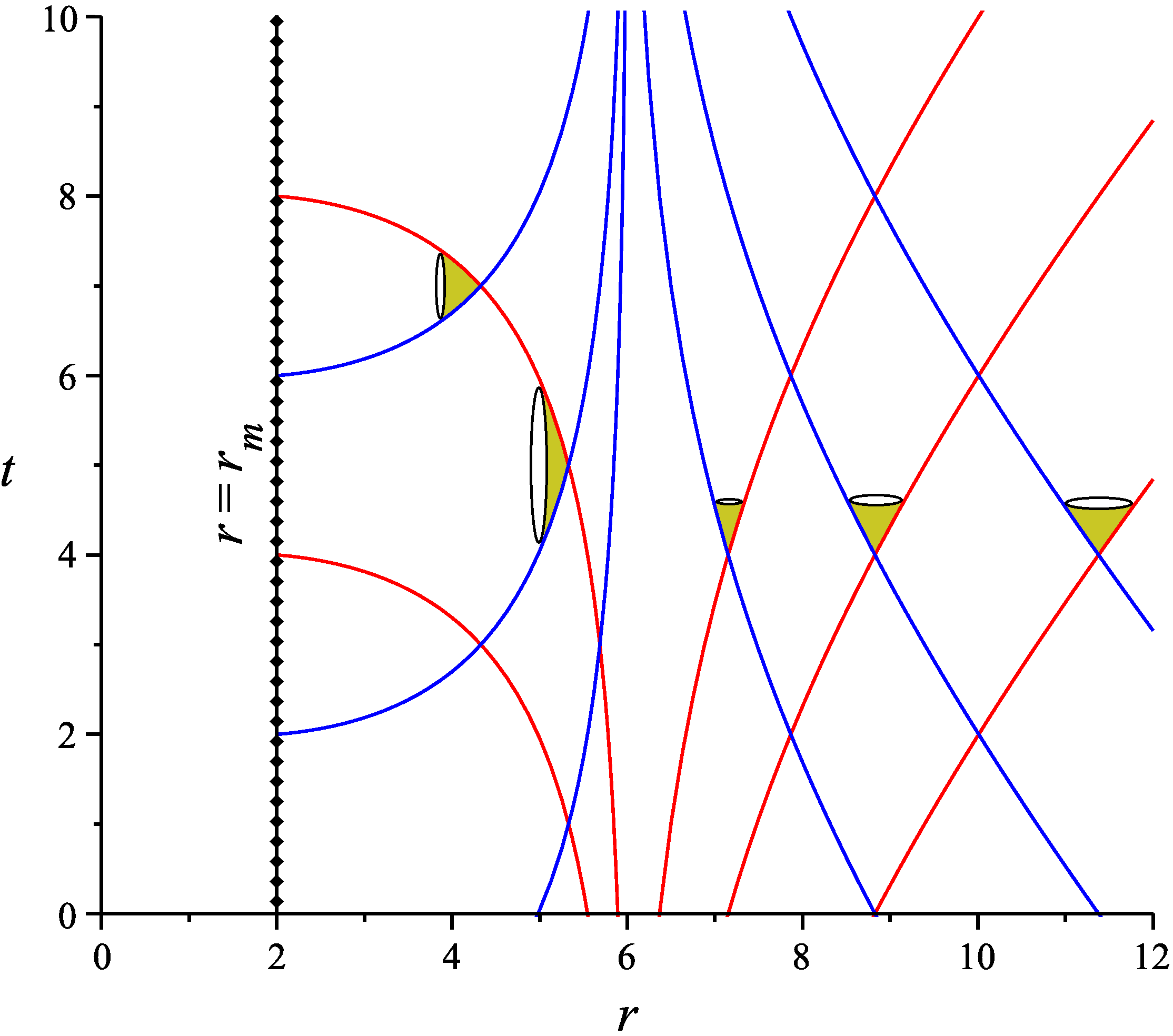}
\centering
\caption{\small{Space-time diagram in Schwarszchild-like coordinates for $r_m=2$ and $r_0=6$.\label{fig4.13}}}
\end{figure}

Let us define a radial coordinate 
\begin{equation}
r^{\ast}=\int\frac{dr}{F(r)} .\label{111}%
\end{equation}
In these coordinates, the null geodesic are given by 
\begin{equation}
t=\pm r^{\ast}+C.\label{112}%
\end{equation}
The figure \ref{fig4.13} shows a space-time diagram with this geodesics.

We can therefore consider $v\equiv t+r^{\ast}$ as a new time
coordinate, which brings the metric on the form 
\begin{equation}
ds^{2}=-F(r)dt^{2}+2dvdr+r^{2}d\Omega_{3}^{2}. \label{113}%
\end{equation}
We now have a non-singular description of particles falling inwards.

Likewise, if we had chosen $u\equiv t-r^{\ast}$ as a new time
coordinate we would have gotten the metric
\begin{equation}
ds^{2}=-F(r)dt^{2}-2dudr+r^{2}d\Omega_{3}^{2}.\label{114}%
\end{equation}
These coordinates have a non-singular description of particles
travelling outwards. 

\begin{figure}[t]
\includegraphics[width=0.6\columnwidth]{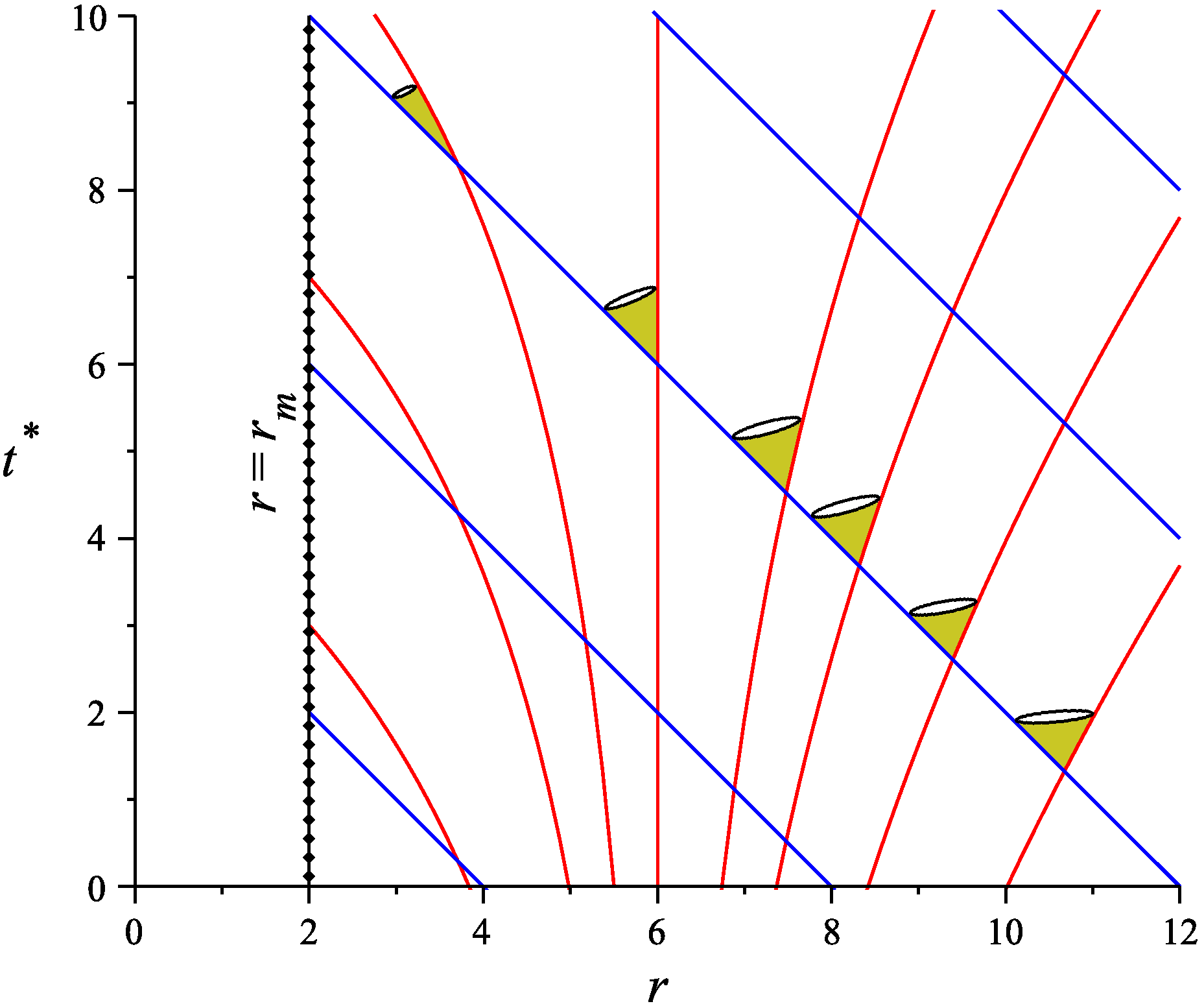}
\centering
\caption{\small{Space-time diagram in advanced Eddington-Finkelstein coordinates.
.\label{fig4.14}}}
\end{figure}

To understand the causal structure in the vicinity of $r=r_{0}$
 is useful to define a new timelike coordinate. In effect let us
define
\begin{equation}
t^{\ast}\equiv v-r, \label{115}%
\end{equation}
so that the ingoing null geodesics are given by
\begin{equation}
t^{\ast}=-r+C \label{116}.%
\end{equation}
These are the straight parallel lines shown on figure \ref{fig4.14}. 

\begin{figure}[t]
\includegraphics[width=0.6\columnwidth]{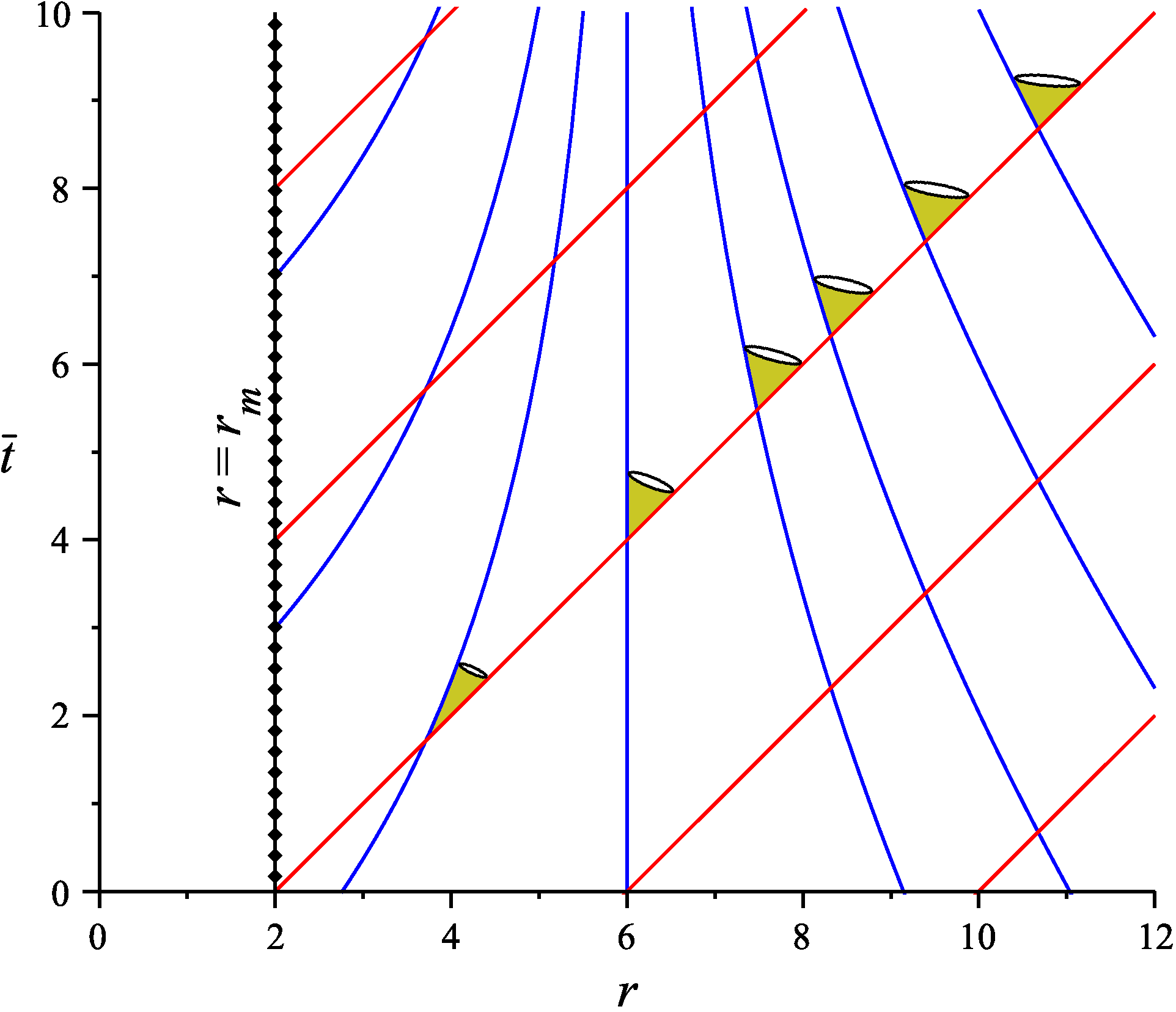}
\centering
\caption{\small{Space-time diagram in retarded Eddington-Finkelstein coordinates
.\label{fig4.15}}}
\end{figure}

On the other hand, it is useful define another timelike coordinate
\begin{equation}
\bar{t}\equiv u+r, \label{116.1}%
\end{equation}
so that the outgoing null geodesics are given by
\begin{equation}
\bar{t}=r+C. \label{116.2}%
\end{equation}
The figure \ref{fig4.15} show a spacetime diagram in these coordinates.

In the coordinate system defined by
\begin{equation}
u =t-r^{\ast},\qquad v  =t+r^{\ast};\qquad -\infty<u,v<\infty,\label{118}%
\end{equation}
the metric is given by
\begin{equation}
ds^{2}=-F(r)dudv+r^{2}d\Omega_{3}^{2}. \label{119}%
\end{equation}

Introducing the Kruskal-Szekeres coordinates 
\begin{equation}
U=-\exp\left(  -\frac{r_{0}}{\sqrt{r_{0}^{4}-r_{m}^{4}}}u\right),\qquad V=\exp\left(  \frac{r_{0}}{\sqrt{r_{0}^{4}-r_{m}^{4}}}v\right) ,
\label{120}%
\end{equation}
we get the result
\begin{equation}
ds^{2}=-\frac{2\sqrt{r_{0}^{4}-r_{m}^{4}}}{r_{0}\exp\left(  \frac{2r_{0}%
}{\sqrt{r_{0}^{4}-r_{m}^{4}}}r^{\ast}(r)\right)  }dUdV+r^{2}d\Omega_{3}%
^{2},\qquad r>r_{0}. \label{121}%
\end{equation}

\begin{figure}[t]
\includegraphics[width=0.6\columnwidth]{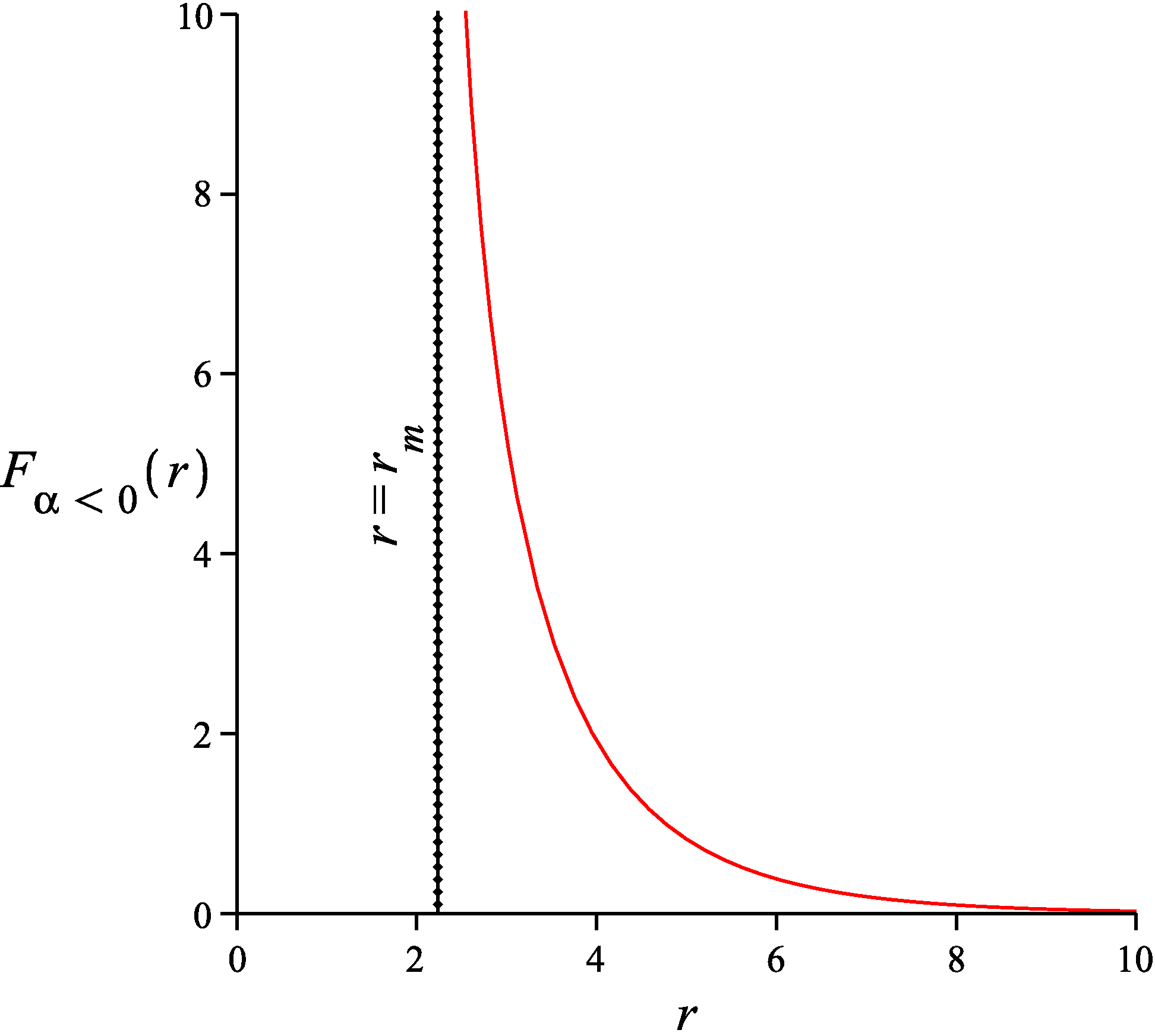}
\centering
\caption{\small{Graph for $F_{\alpha<0}(r)$ with $l=1$, $r_m=\sqrt{5}$ and $r_0=\sqrt{13}$.\label{fig4.16}}}
\end{figure}

Defining the function
\begin{equation}
F_{\alpha<0}(r)=\frac{2\sqrt{r_{0}^{4}-r_{m}^{4}} F(r)
}{r_{0}\exp\left(  \frac{2r_{0}}{\sqrt{r_{0}^{4}-r_{m}^{4}}}r^{\ast
}(r)\right)  },\qquad r>r_{m},\label{122}%
\end{equation}
whose graph is shown in figure \ref{fig4.16}.

Therefore, we have
\begin{equation}
ds^{2}=-F_{\alpha<0}(r)dUdV+r^{2}d\Omega_{3}^{2},\quad r>r_{m}.
\end{equation}

Using the Kruskal coordinates 
\begin{equation}
T=\frac{1}{2}(U+V),\qquad X=-\frac{1}{2}(U-V) \label{123},
\end{equation}
we can see that the line element takes the form 
\begin{equation}
ds^{2}=-F_{\alpha}(r)\left(  -dT^{2}+dX^{2}\right)  +r^{2}d\Omega_{3}^{2}.
\label{124}%
\end{equation}

\begin{figure}[t]
\includegraphics[width=0.6\columnwidth]{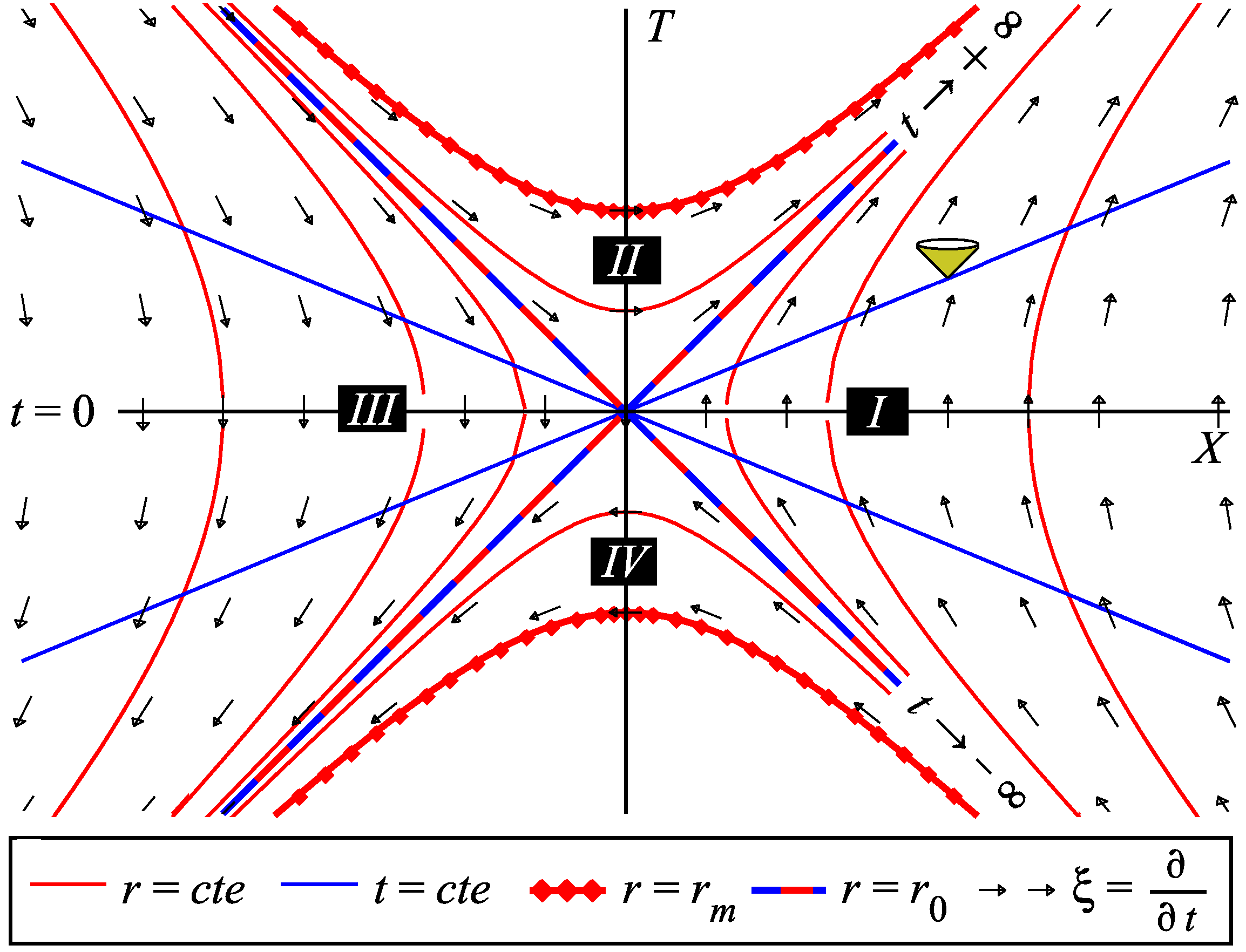}
\centering
\caption{Kruskal-Szekeres diagram.\small{\label{fig4.17}}}
\end{figure}

In figure \ref{fig4.17} we have illustrated the corresponding
Kruskal-Szekeres diagram.

We conducted the compactification, defining the following null coordinate $q$
and $p$
\begin{equation}
U   =\tan q,\qquad V =\tan p .
\end{equation}

Now we introduce the coordinates defined by
\begin{equation}
\tau=p+q\quad\text{and}\quad x=p-q .
\end{equation}

\begin{figure}[t]
\includegraphics[width=0.6\columnwidth]{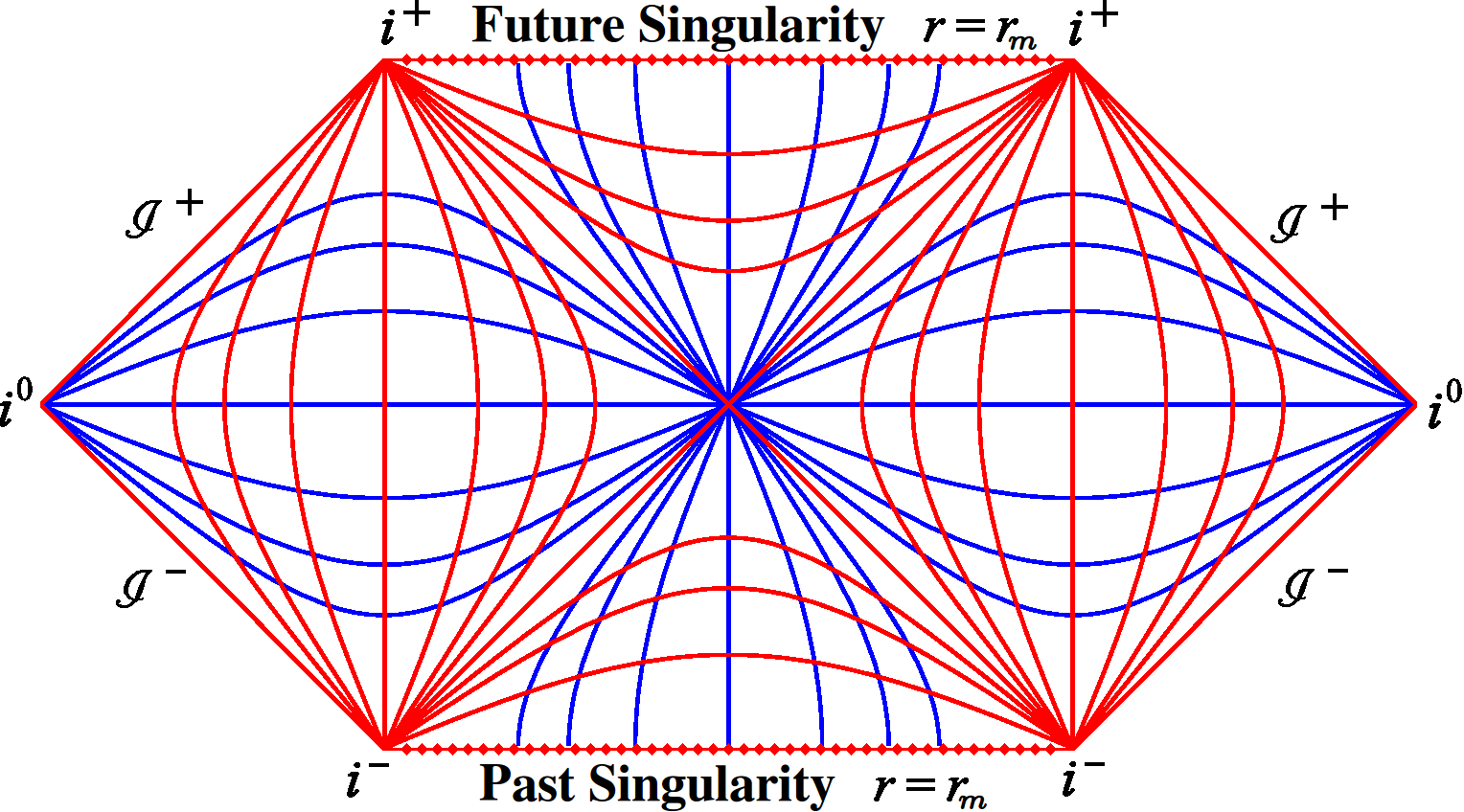}
\centering
\caption{\small{Penrose diagram for $\alpha<0$ with $\frac{\kappa_{E}}{6\pi^{2}}M>l^{2}$.\label{fig4.18}}}
\end{figure}

The figure \ref{fig4.18} shows the corresponding Penrose diagram.

\subsection{Case $\alpha<0$: Naked singularity}

In this section we considerer that
\begin{equation}
\frac{\kappa_{EC}}{6\pi^2}<l^2,
\end{equation}
or equivalently
\begin{equation}
r_m<l.
\end{equation}

Therefore we have
\begin{equation}
F=1-\frac{r^2}{l^2}+\sqrt{\frac{r^4-r_m^4}{l^4}},
\end{equation}
this function have no real roots. This mean that the metric is not singular at $r\neq r_m$.

\begin{figure}[t]
\includegraphics[width=0.6\columnwidth]{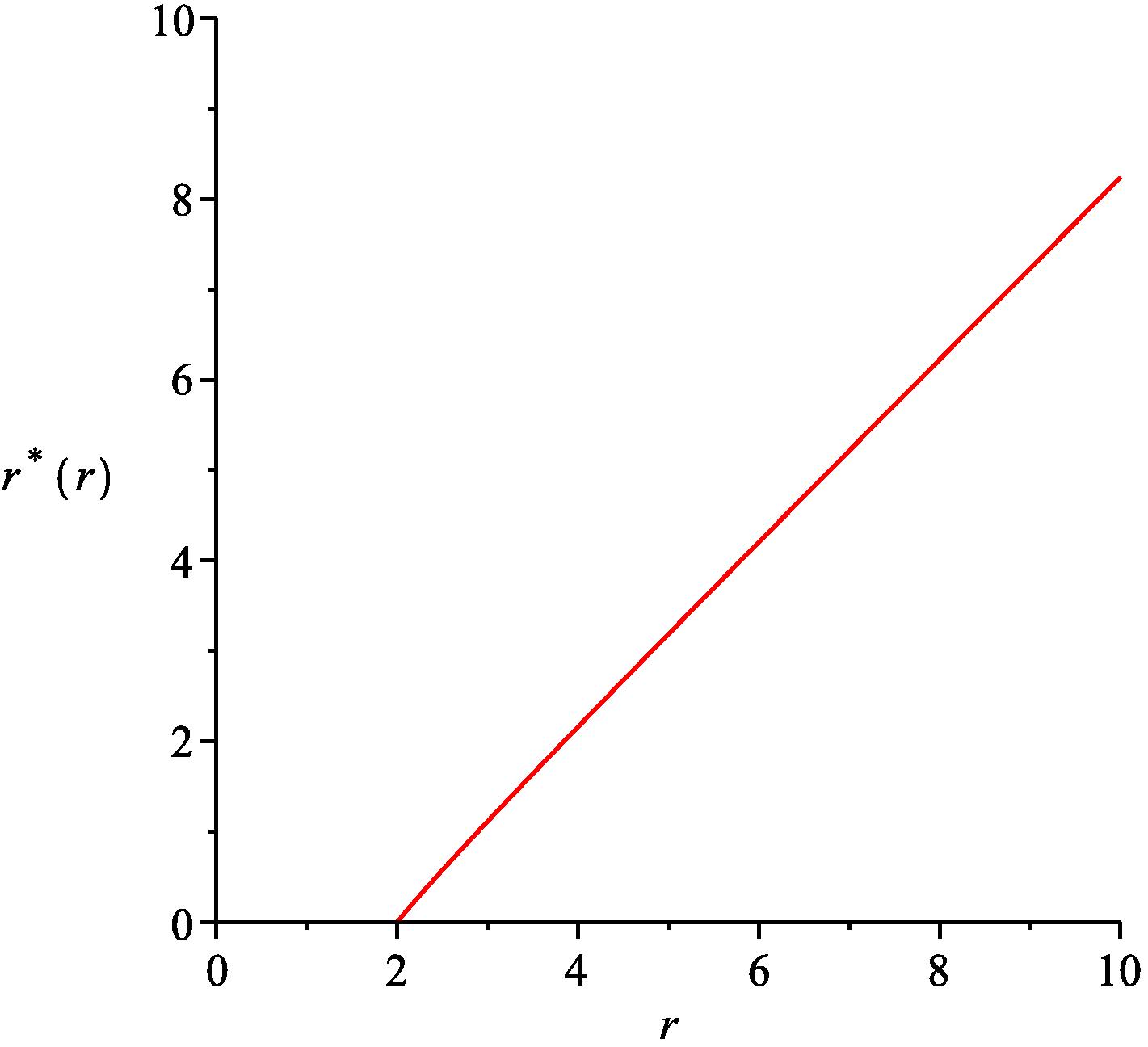}
\centering
\caption{\small{Graph for $r^*(r)$ with $\frac{\kappa_{EC}}{6\pi^2}M=1$ and $l=4$, so that $r_m=2$ and $r_0=\sqrt{\frac{17}{2}}\approx 2.92$ .\label{fig4.19}}}
\end{figure}

Let us define a new radial coordinate
\begin{equation}
r^*(r)=\int \frac{dr}{F(r)},
\end{equation}
with $r^*(r_m)=0$, we obtain (see appendix \ref{int04})
\begin{align}
r^{*} =\ & \frac{r-r_m}{2}+\frac{\sqrt{r_0^4-r_m^4}}{4r_0}\Biggl(\ln\left(\frac{r+r_0}{r_m+r_0}\right)+\mathrm{Z}_{\alpha<0}(r)\Biggr)\nonumber\\
&+\frac{r_m}{2}\left\{\mathrm F\left(i\frac{r}{r_m},i\right)-\mathrm E\left(i\frac{r}{r_m},i\right)-\mathrm F\left(i,i\right)+\mathrm E\left(i,i\right)\right\}\\
& +\frac{r_0^2}{2r_m}\left\{\mathrm F\left(i\frac{r}{r_m},i\right)-\mathrm F\left(i,i\right)\right\},\nonumber
\end{align}
where $r_0=\sqrt{\frac{\kappa_{E}}{12\pi^2}M+\frac{l^2}{2}}$. A graph for $r^*(r)$ is shown in the figure \ref{fig4.19}.

In these coordinates, the null geodesics are given by
\begin{equation}
t=\pm r^*(r)+C
\end{equation}

Since no singularities can be removed ($r=r_m$ is a singularity of the space-time), we have no maximal extensions for this solution. Consider then the corresponding conformal compactification. For this purpose we can define the radial null coordinates, incoming and outgoing 
\begin{equation}
u=t-r^*\qquad\text{and}\qquad  v=t+r^*.
\end{equation}

The Kruskal-type corrdinates are defined by
\begin{equation}
U=-\exp\left(-\frac{u}{2}\right),\qquad V=\exp\left(\frac{v}{2}\right).
\end{equation}

\begin{figure}[t]
\includegraphics[width=0.25\columnwidth]{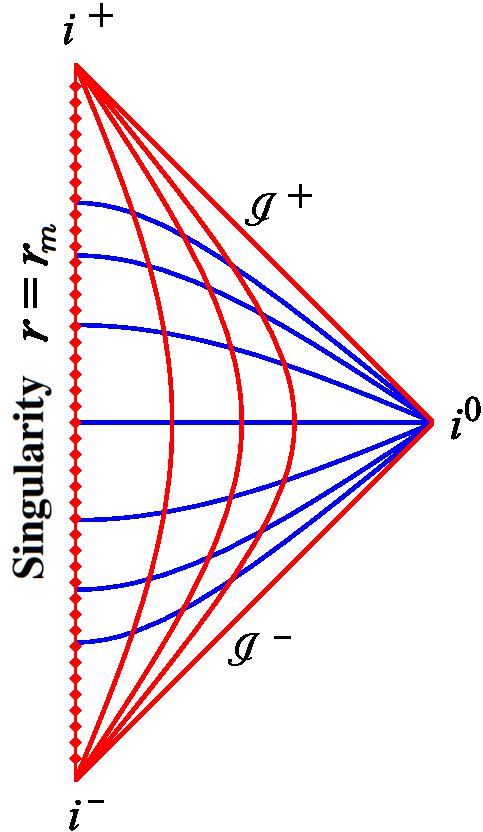}
\centering
\caption{\small{Penrose diagram for $\alpha<0$ and $
\frac{\kappa_{E}}{6\pi^{2}}M<l^{2}$ .\label{fig4.20}}}
\end{figure}

Then, we conducted the compactification defining another radial null coordinates given by
\begin{equation}
U=\tan q,\qquad  V=\tan q,
\end{equation}
so that we introduce a time-like and a space-like coordinates ($\tau-x$) with which we construct the Penrose diagram given in figure \ref{fig4.20}.

\section{Summary and Outlook}

We have considered a 5-dimensional action $S=S_{g}+S_{M}$ which is composed of
a gravitational sector and a sector of matter, where the gravitational sector
is given by a Chern-Simons gravity action instead of the Einstein-Hilbert
action.  We studied the implications that has on the Black Holes solutions,
the fact of replacing the Einstein-Hilbert lagrangian by the Chern-Simons
lagrangian in the gravitational sector of the action. 

We have found some solutions for the Einstein-Chern-Simons field equations,
which were obtained from the action $S=S_{g}+S_{M}$ where $S_{g}$ is the
action for the Einstein-Chern-Simons gravity theory, studied in Ref.
\cite{salg1}.

\acknowledgments 
This work was supported in part by Direcci\'{o}n de Investigaci\'{o}n,
Universidad de Concepci\'{o}n through Grant \# 210.011.053-1.0 and in part by
FONDECYT through Grants \# 1080530.

\appendix

\section{Static and spherically symmetric solutions for Eintein-Cartan gravity in 5D}

We consider the Einstein field equation
\begin{equation}
\varepsilon_{abcde}R^{bc}e^{d}e^{e}  =-\frac{\delta L_{M}}{\delta e^{a}%
},\qquad T^{\text{ }a}=0\nonumber\\
\end{equation}
we take the Hodge dual on first equation
\begin{equation}
X_{a}=\kappa_E T_{ab}e^{b} \label{A1}%
\end{equation}
where 
\begin{equation}\label{A1.1}%
X_{a}=\star\left(  \varepsilon_{abcde}R^{bc}e^{d}e^{e}\right)
\end{equation}
and $T_{ab}$ is the energy-momentum tensor of the matter.

In five dimensions the spherically-and static- symmetric metric is given by%
\begin{equation}
ds^{2}=-e^{2f(r)}dt^{2}+e^{2g(r)}dr^{2}+r^{2}d\Omega_{3}^{2} \label{A2}%
\end{equation}%
where
\begin{equation}
d\Omega_{3}^{2}=d\theta_{1}^{2}+\sin^{2}\theta_{1}d\theta_{2}^{2}+\sin
^{2}\theta_{1}\sin^{2}\theta_{2}d\theta_{3}^{2} \label{A3}%
\end{equation}

Introducing an orthonormal basis ($ds^2=\eta_{ab}e^ae^b$):
\begin{equation*}
\eta_{ab}=\mathrm{diag}(-1,+1,+1,+1,+1),
\end{equation*}
\begin{align}
e^{T}=e^{f(r)}dt,\quad e^{R}=e^{g(r)}dr,\quad e^{1}=rd\theta_{1},\quad e^{2}=r\sin\theta_{1}d\theta_{2},\quad e^{3}=r\sin\theta_{1}\sin\theta_{2}d\theta_{3}. \label{A4}%
\end{align}

We can use Cartan's first structural equation ($T^{a}=de^{a}+\omega_{\text{ }b}^{a}e^{b}=0$) and the antisymmetry of the connection forms, $\omega^{ab}=-\omega^{ba}$, to
find the non-zero connection forms. The calculations give:%
\begin{align}
\omega_{\text{ }TR}&=-f^{\prime}e^{-g}e^{T},\qquad\omega_{Ri}=-\frac{e^{-g}}{r}e^{i},\qquad\omega_{12}=-\frac{1}{r\tan\theta_{1}}e^{2},\nonumber\\
\omega_{13}&=-\frac{1}{r\tan\theta_{1}}e^{3},\qquad
\omega_{23}=-\frac{1}{r\sin\theta_{1}\tan\theta_{2}}e^{3};\qquad i=1,2,3.\label{A7}
\end{align}

From Cartan's second structural equation ($R_{\phantom{a} b}^{a}=d\omega_{\phantom{a}b}^{a}+\omega_{\phantom{a} c}^{a}\omega_{\phantom{a} b}^{c}$)
we can calculate the curvature matrix. The non-zero components are%
\begin{align}
R^{TR}&=e^{-g}\left(  f^{\prime}g^{\prime}-f^{\prime\prime}-\left(  f^{\prime
}\right)  ^{2}\right)  e^{T}e^{R},\qquad R^{Ti}=-\frac{f^{\prime}e^{-2g}}{r}e^{T}e^{i}\nonumber\\
R^{Ri}&=\frac{g^{\prime}e^{-2g}}{r}e^{R}e^{i},\qquad R^{ij}=\frac{1-e^{-2g}}{r^{2}}e^{i}e^{j};\qquad i,j=1,2,3 .\label{A9}
\end{align}

Introducing (\ref{A4}), (\ref{A9}) into (\ref{A1.1})
\begin{align}
X_{T}  &  =12\frac{e^{-2g}}{r^{2}}\left(  g^{\prime}r+e^{2g}-1\right)
e^{T},\nonumber\\
X_{R}  &  =12\frac{e^{-2g}}{r^{2}}\left(  f^{\prime}r-e^{2g}+1\right)
e^{R},\label{A10}\\
X_{i}  &  =4\frac{e^{-2g}}{r^{2}}\left(  -f^{\prime}g^{\prime}r^{2}%
+f^{\prime\prime}r^{2}+\left(  f^{\prime}\right)  ^{2}r^{2}+2f^{\prime
}r-2g^{\prime}r-e^{2g}+1\right)  e^{i};\text{ \ }i=1,2,3.\nonumber
\end{align}

Introducing (\ref{A10}) into (\ref{A1}) and considering the energy-momentum
tensor as the energy-momentum tensor of a perfect fluid at rest, i.e., $T_{TT}%
=\rho(r)$ and $T_{RR}=T_{ii}=P(r),$ where $\rho(r)$ and $P(r)$ are the energy
density and pressure we find
\begin{align}
12\frac{e^{-2g}}{r^{2}}\left(  rg^{\prime}+e^{2g}-1\right) &=\kappa_E
\rho, \label{A11}\\
12\frac{e^{-2g}}{r^{2}}\left(  rf^{\prime}-e^{2g}+1\right)  &=\kappa_E P,
\label{A12}\\
4\frac{e^{-2g}}{r^{2}}\left(  -f^{\prime}g^{\prime}r^{2}+f^{\prime\prime}%
r^{2}+\left(  f^{\prime}\right)  ^{2}r^{2}+2f^{\prime}r-2g^{\prime}%
r-e^{2g}+1\right)  &=\kappa_E P. \label{A13}%
\end{align}

\subsection{The Exterior Solution}

If $\rho(r)=P(r)=0$ the field equation are given by
\begin{align}
12\frac{e^{-2g}}{r^{2}}\left(  rg^{\prime}+e^{2g}-1\right) &=0, \label{A14}\\
12\frac{e^{-2g}}{r^{2}}\left(  rf^{\prime}-e^{2g}+1\right)  &=0,
\label{A15}\\
4\frac{e^{-2g}}{r^{2}}\left(  -f^{\prime}g^{\prime}r^{2}+f^{\prime\prime}%
r^{2}+\left(  f^{\prime}\right)  ^{2}r^{2}+2f^{\prime}r-2g^{\prime}%
r-e^{2g}+1\right)  &=0. \label{A16}%
\end{align}

Consider the equation (\ref{A14}). After multiplying (\ref{A14}) by
$r^{3}/6$ we find
\begin{equation}
\Bigl( r^{2}\left(  1-e^{-2g}\right)  \Bigr)  ^{\prime}=0. \label{A17}%
\end{equation}
Integrating we have
\begin{equation}
e^{-2g}=1-\frac{\kappa_E }{12\pi^2 r^{2}}M.\label{A18}%
\end{equation}

Adding equations (\ref{A15}) and (\ref{A16}) we find
\begin{equation}
e^{2f}=e^{-2g}=1-\frac{\kappa_{E}M}{12\pi^{2}r^{2}}, \label{A19}%
\end{equation}
and equation (\ref{A16}) is satisfied.

\subsection{The Interior Solution}
Now consider the equation (\ref{A11}). After multiplying by $r^{3}/6$ we find%
\begin{equation}
\Bigl(  r^{2}\left(  1-e^{-2g}\right)  \Bigr)  ^{\prime}=\frac{\kappa}%
{6}\rho r^{3}. \label{A20}%
\end{equation}
Integrating we have
\begin{equation}
r^{2}\left(  1-e^{-2g}\right)  =\frac{\kappa}{12\pi^{2}}\Bigl( \mathcal{M}(r)-\mathcal{M}_{0}%
\Bigr) , \label{A21}%
\end{equation}
where $\mathcal{M}_{0}$ is an integration constant and $\mathcal{M}(r)$ is the Newtonian mass,
which is defined as%
\begin{equation}
\mathcal{M}(r)=2\pi^{2}\int_{0}^{r}\rho(\bar{r})\bar{r}^3d\bar{r}, \label{A22}%
\end{equation}
so that
\begin{equation}
e^{-2g}=1-\frac{\kappa}{12\pi^{2}r^2}\Bigl( \mathcal{M}(r)-\mathcal{M}_{0}\Bigr).
\label{A23}%
\end{equation}
To eliminate the singularity at $r=0$ put $\mathcal{M}_{0}=0,$ then%
\begin{equation}
e^{-2g}=1-\frac{\kappa}{12\pi^{2}r^{2}}\mathcal{M}(r). \label{A24}%
\end{equation}

\subsection{The Tolman-Oppenheimer-Volkoff equation in 5D}

Our interest is to compute the pressure and density of matter in a spherically
symmetric, static star. Since we are assuming spherical symmetry the metric
will be of the form (\ref{A2}). Let us recall that the energy-momentum
tensor satisfies the condition %
\begin{equation}
\nabla_{\mu}T^{\mu\nu}=0. \label{A25}%
\end{equation}

If $T_{TT}=\rho(r)$ and $T_{RR}=T_{ii}=P(r)$ we find
\[
\nabla_{\mu}T^{\mu r}=\frac{f^{\prime}\Bigl( \rho(r)+P(r)\Bigr)  +P^{\prime
}(r)}{e^{2g}}=0,
\]
so that
\begin{equation}
f^{\prime}=-\frac{P^{\prime}}{\rho+P}, \label{A26}%
\end{equation}
expression known as \emph{hydrostatic equilibrium equation}.

From (\ref{A12}) and (\ref{A24}) we find
\begin{equation}
f^{\prime}(r)=\frac{\kappa_E\mathcal{M}(r)}{12\pi^{2}r^{3}}\left(  1+\frac{\pi^{2}%
r^{4}P(r)}{\mathcal{M}(r)}\right)  \biggl(  1-\frac{\kappa}{12\pi^{2}r^{2}}\mathcal{M}(r)\biggr)
^{-1}. \label{A27}%
\end{equation}
Introducing (\ref{A26}) into (\ref{A27}) we obtain the following equation%
\begin{equation}
P^{\prime}(r)=-\frac{\kappa_E\mathcal{M}(r)}{12\pi^{2}r^{3}}\left(  1+\frac
{P(r)}{\rho(r)}\right)  \left(  1+\frac{\pi^{2}%
r^{4}P(r)}{\mathcal{M}(r)}\right)  \biggl(  1-\frac{\kappa}{12\pi^{2}r^{2}}\mathcal{M}(r)\biggr)
^{-1} ,\label{A28}%
\end{equation}
which is the five-dimensional \emph{Tolman-Oppenheimer-Volkoff} equation. Compare with the 4-dimensional case shown in equation (1.11.13) of the reference \cite{weinberg}.

We can resolve this equation for objects that are isentropic,
that is, in which the entropy per nucleon does not vary throughout the it. For example, we have two very different kinds of star that satisfies this condition: (i) \emph{stars at absolute zero}. According to Nernst's theorem, the entropy per nucleon will then be zero throughout the star and (ii) \emph{stars in convective equilibrium}. If the most efficient mechanism for energy
transfer within the star is convection, then in equilibrium the entropy per nucleon
must be nearly constant throughout the star. We also assume that the stars we consider have a chemical composition that is constant throughout.

With preceding assumptions, the pressure $P$ may be expressed as a function of the density $\rho$, the entropy per nucleon $s$, and the chemical composition. So, with $s$ and the chemical composition constant throughout the star, $P(r)$ may be regarded as a function of $\rho(r)$.

Given an equation of state $P(\rho)$, we now formulate our problem as a pair of
first-order differential equations for $P(r)$, $\mathcal{M}(r)$ and $\rho(r)$, the equation (\ref{A28}) and 
\begin{equation}
\mathcal{M}^{\prime}(r)=2\pi^{2}r^3\rho(r), \label{A22'}%
\end{equation}
with an initial condition $\mathcal{M}(0)=0$. In addition, it is necesary to provide other initial condition, that is, the value $\rho(0)=\rho_0$.

The differential equations must be integrated out from the center of the star, until
$P(\rho(r))$ drops to zero at some point $r = R$, which we then interpret as the radius
of star.

Let us return to the problem of calculating the metric. Once we compute
$\rho(r)$, $\mathcal{M}(r)$ and $P(r)$, we can immediately obtain $g(r)$ from equation (\ref{A24}) and $f(r)$ from the equation (\ref{A27})
\begin{equation}
f(r)=-\int_r^\infty\frac{\kappa_E\mathcal{M}(\bar{r})}{12\pi^{2}\bar{r}^{3}}\left(  1+\frac{\pi^{2}%
\bar{r}^{4}P(\bar{r})}{\mathcal{M}(\bar{r})}\right)  \biggl(  1-\frac{\kappa}{12\pi^{2}\bar{r}^{2}}\mathcal{M}(\bar{r})\biggr)
^{-1}\ d\bar{r}, \label{A27'}%
\end{equation}
where we have set $f(\infty)=0$, condition consistent with the asymptotic limit from the exterior solution.

\section{Dynamics of the field $h^{a}$}
So far we have interpreted the field $h^{a}$ as a field of matter whose nature
has not been specified.

We consider now the field $h^{a}$. Expanding the field $h^{a}$ in their holonomic index we have%
\begin{equation}
h^{a}=h_{b\nu}\ \eta^{ab}\ dx^{\nu}=h_{\mu\nu}\ \eta^{ab}\ e_b^\mu \ dx^{\nu}.
\end{equation}

Whether the space-time is static and spherically symmetric, the field $h_{\mu\nu}$ therefore must satisfy the Killing equation $\mathcal{L}_\xi h_{\mu\nu}=0$ for $\xi_0=\partial_t$ (stationary) and the six generators of the sphere $S^3$
\begin{align}
&\xi_1=\partial_{\theta_3},\qquad \xi_2=\sin\theta_3\ \partial_{\theta_2}+\cot\theta_2\cos\theta_3\ \partial_{\theta_3}\nonumber\\
&\xi_3=\cos\theta_3\ \partial_{\theta_2}-\cot\theta_2\sin\theta_3\ \partial_{\theta_3},\qquad \xi_4=\cos\theta_2\ \partial_{\theta_1}-\cot\theta_1\sin\theta_2\ \partial_{\theta_2}\nonumber\\
&\xi_5=\sin\theta_2\sin\theta_3\ \partial_{\theta_1}+\cot\theta_1\cos\theta_2\sin\theta_3\ \partial_{\theta_2}+\cot\theta_1\csc\theta_2\cos\theta_3\ \partial_{\theta_3}\nonumber\\
&\xi_6=\sin\theta_2\cos\theta_3\ \partial_{\theta_1}+\cot\theta_1\cos\theta_2\cos\theta_3\ \partial_{\theta_2}-\cot\theta_1\csc\theta_2\sin\theta_3\ \partial_{\theta_3}.
\end{align}

Then, we have%
\begin{align}
h^T&=h_{t}(r)\ e^T+h_{tr}(r)\ e^R\nonumber\\
h^R&=h_{rt}(r)\ e^T+h_{r}(r)\ e^R\nonumber\\
h^i&= h(r)\ e^i
\end{align}

From (\ref{treintay4}) and (\ref{33}) and replacing in the second field equation from (\ref{30-1}), we can see that%
\begin{equation}
h_{tr}=h_{rt}=0,
\end{equation}
and
\begin{equation}
h_r=(rh)',\qquad h'_t=f'(h_r-h_t).
\end{equation}

\section{Integrals}\label{int00}
\subsection{Elliptic Integrals}
The incomplete elliptic integral of the first kind is defined as

\begin{equation}
\mathrm{F}(z,k):=\int_0^z \frac{dt}{\sqrt{1-t^2}\sqrt{1-k^2t^2}}.
\end{equation}

The incomplete elliptic integral of the second kind is defined as

\begin{equation}
\mathrm{E}(z,k):=\int_0^z \frac{\sqrt{1-k^2t^2}}{\sqrt{1-t^2}}dt.
\end{equation}

The incomplete elliptic integral of the third kind is defined as

\begin{equation}
\Pi(z,v,k):=\int_0^z \frac{dt}{(1-vt^2)\sqrt{1-t^2}\sqrt{1-k^2t^2}}.
\end{equation}

For more information you can see ref. \cite{abramo}.

\subsection{Case $\alpha>0$ and $\frac{\kappa_{E}}{6\pi^2}M>l^2$}\label{int01}

\begin{equation}
r^*(r)=\int\frac{dr}{ 1+\frac{r^2}{l^2}+\sqrt{\frac{r^4+2r_0^2l^2+l^4}{l^4}}},
\end{equation}
where $r_0=\sqrt{\frac{\kappa_{E}}{12\pi^2}M-\frac{l^2}{2}}$ is that 
\begin{equation}
1+\frac{r^2}{l^2}+\sqrt{\frac{r^4+2r_0^2l^2+l^4}{l^4}}\Bigg|_{r=r_0}=0.
\end{equation}

We can separate $r^*$ in the following way
\begin{equation}
r^*=\frac{r}{2}+\frac{r_0^2+l^2}{2} I_{11}+\frac{1}{2}I_{21}+\frac{r_0^2}{2}I_{22}+\frac{\left(r_0^2+l^2\right)^2}{2}I_{23},
\end{equation}
where
\begin{align}
I_{11}(r)=&\ \int\frac{dr}{r^2-r_0^2}\quad ,\quad I_{21}(r)=\int\frac{r^2}{\sqrt{r^4+2r_0^2l^2+l^4}}\ dr\nonumber\\
I_{22}(r)=&\ \int\frac{dr}{\sqrt{r^4+2r_0^2l^2+l^4}}\quad ,\quad  I_{23}(r)=\int\frac{dr}{(r^2-r_0^2)\sqrt{r^4+2r_0^2l^2+l^4}}
\end{align}
the computations are
\begin{align}
I_{11}(r)=&\frac{1}{2r_0}\ln\left|\frac{r-r_0}{r+r_0}\right|\nonumber\\
I_{21}(r)=&\sqrt{il\sqrt{2r_0^2+l^2}}\Biggl\{ \mathrm{F}\left(\sqrt{\frac{i}{l\sqrt{2r_0^2+l^2}}}\ r,i\right)-\mathrm{E}\left(\sqrt{\frac{i}{l\sqrt{2r_0^2+l^2}}}\ r,i\right)\Biggr\}\nonumber\\
I_{22}(r)=&-i\sqrt{\frac{i}{l\sqrt{2r_0^2+l^2}}}\ \mathrm{F}\left(\sqrt{\frac{i}{l\sqrt{2r_0^2+l^2}}}\ r,i\right)\\
I_{23}(r)=&\frac{i}{r_0^2l}\sqrt{\frac{il}{\sqrt{2r_0^2+l^2}}}\ \Pi\left(\sqrt{\frac{i}{l\sqrt{2r_0^2+l^2}}}\ r,-\frac{il\sqrt{2r_0^2+l^2}}{r_0^2},i\right)\nonumber
\end{align}

However $I_{23}$ has been computed with help from the incomplete elliptic integral of the third kind. Sadly, this way cannot get the correct result for input data provided. To solve this problem, we can separate the non finite part from the integrand, we obtain
\begin{align}
\frac{1}{(r^2-r_0^2)\sqrt{r^4+2r_0^2l^2+l^4}}=&\frac{1}{2r_0(r_0^2+l^2)(r-r_0)}\nonumber\\
&\ +\frac{2r_0(r_0^2+l^2)-(r+r_0)\sqrt{r^4+2r_0^2l^2+l^4}}{2r_0(r_0^2+l^2)(r^2-r_0^2)\sqrt{r^4+2r_0^2l^2+l^4}},
\end{align}
so that we can immediately integrate to obtain
\begin{equation}
I_{23}=\frac{1}{2r_0(r_0^2+l^2)}\left(\ln\left|\frac{r-r_0}{r_0}\right|+\mathrm{Z}_{\alpha>0}(r)\right),
\end{equation}
where we define the following smooth function to be computed through numerical methods
\begin{equation}
\mathrm{Z}_{\alpha>0}(r)=\int_0^r\frac{2r_0(r_0^2+l^2)-(t+r_0)\sqrt{t^4+2r_0^2l^2+l^4}}{(t^2-r_0^2)\sqrt{t^4+2r_0^2l^2+l^4}}\ dr.
\end{equation}

Collecting all contributions and setting $r^*(0)=0$
\begin{align}
r^*(r)=&\ \frac{r}{2}+\frac{r_0^2+l^2}{4r_0}\Biggl\{\ln\left(\frac{(r-r_0)^2}{r_0(r+r_0)}\right)+\mathrm{Z}_{\alpha>0}(r)\Biggr\}\nonumber\\
&-\frac{ir_0^2}{2}\sqrt{\frac{i}{l\sqrt{2r_0^2+l^2}}}\ \mathrm{F}\left(\sqrt{\frac{i}{l\sqrt{2r_0^2+l^2}}}\ r,i\right)\\
&+\frac{1}{2}\sqrt{il\sqrt{2r_0^2+l^2}}\Biggl\{ \mathrm{F}\left(\sqrt{\frac{i}{l\sqrt{2r_0^2+l^2}}}\ r,i\right)-\mathrm{E}\left(\sqrt{\frac{i}{l\sqrt{2r_0^2+l^2}}}\ r,i\right)\Biggr\}.\nonumber
\end{align}

\subsection{Case $\alpha>0$ and $\frac{\kappa_{E}}{6\pi^2}M<l^2$}\label{int02}
Now we will compute
\begin{equation}
r^*(r)=\int\frac{1}{1+\frac{r^2}{l^2}-\sqrt{\frac{r^4+ml^2}{l^4}}}\ dr,
\end{equation}
where $m=\frac{\kappa_{EC}}{6\pi^2}M$ and $m<l^2$. It is important to note that
\begin{equation}
1+\frac{r^2}{l^2}-\sqrt{\frac{r^4+ml^2}{l^4}}\neq 0\quad,\quad \forall r >0.
\end{equation}

We can separate $r^∗$ in the following way
\begin{equation}
r^*=\frac{r}{2}+\frac{l^2+m}{4}I_{11}+\frac{1}{2}I_{21}-\frac{l^2-m}{4}I_{22}+\frac{\left(l^2+m\right)^2}{8}I_{23},
\end{equation}
where
\begin{align}
I_{11}=&\ \int\frac{dr}{r^2+\frac{l^2-m}{2}}\quad,\quad I_{21}=\int\frac{r^2}{\sqrt{r^4+ml^2}}\ dr\\
I_{22}=&\ \int\frac{dr}{\sqrt{r^4+ml^2}}\quad,\quad I_{23}=\int\frac{dr}{\left(r^2-\frac{l^2-m}{2}\right)\sqrt{r^4+ml^2}},\nonumber
\end{align}
whose results are
\begin{align}
I_{11}(r)=&\ \sqrt{\frac{2}{l^2-m}}\arctan\left(\sqrt{\frac{2}{l^2-m}}\ r\right)\nonumber\\
I_{21}(r)=&\ \sqrt{il\sqrt{m}}\left\{ \mathrm{F}\left(
\sqrt{\frac{i}{l\sqrt{m}}}\ r,i\right)  -\mathrm{E}\left(  \sqrt{\frac
{i}{l\sqrt{m}}}\ r,i\right)  \right\} \nonumber\\
I_{22}(r)=&\ -i\sqrt{\frac{i}{l\sqrt{m}}}\ \mathrm{F}\left(  \sqrt
{\frac{i}{l\sqrt{m}}}\ r,i\right)\\
I_{23}(r)=&\  -\frac{2i}{l^2-m}\sqrt{\frac
{i}{l\sqrt{m}}}\ \Pi\left(  \sqrt{\frac{i}{l\sqrt{m}}}r,\frac{2i\ l\sqrt{m}}%
{l^2-m},i\right).\nonumber
\end{align}
This time, the incomplete elliptic integral of the third kind have no problem.

Collecting all contributions and setting $r^*(0)=0$.
\begin{align}
r^{\ast}\left(  r\right)   &  =\frac{r}{2}+\frac{\sqrt{2}}{4}\frac{l^2+m}{\sqrt{(l^2-m)}}%
\arctan\left(  \sqrt{\frac{2}{l^2-m}}\ r\right) \nonumber\\
&  +\frac{1}{2}\sqrt{il\sqrt{m}}\left\{ \mathrm{F}\left(
\sqrt{\frac{i}{l\sqrt{m}}}\ r,i\right)  -\mathrm{E}\left(  \sqrt{\frac
{i}{l\sqrt{m}}}\ r,i\right)  \right\} \label{98'}\\
&  +\frac{i(l^2-m)}{4}\sqrt{\frac{i}{l\sqrt{m}}}\ \mathrm{F}\left(  \sqrt
{\frac{i}{l\sqrt{m}}}\ r,i\right)  -\frac{i\left(l^2+m\right)^{2}}{4(l^2-m)}\sqrt{\frac
{i}{l\sqrt{m}}}\Pi\left(  \sqrt{\frac{i}{l\sqrt{m}}}r,\frac{2il\sqrt{m}}%
{l^2-m},i\right) \nonumber
\end{align}

\subsection{Case $\alpha<0$ and $\frac{\kappa_{E}}{6\pi^2}M>l^2$}\label{int03}
The next integral to be computing is
\begin{equation}
r^*(r)=\int\frac{1}{1-\frac{r^2}{l^2}+\sqrt{\frac{r^4-2l^2r_0^2+l^4}{l^4}}}\ dr,
\end{equation}
where $r_0=\sqrt{\frac{\kappa_{E}}{12\pi^2}M+\frac{l^2}{2}}$ is that
\begin{equation}
1-\frac{r^2}{l^2}+\sqrt{\frac{r^4-2l^2r_0^2+l^4}{l^4}}\Bigg|_{r=r_0}=0
\end{equation}

You can note there is a minimum value for $r$ given by $r_m=\sqrt[4]{l^2(2r_0^2-l^2)}=\sqrt[4]{\frac{\kappa_{E}}{6\pi^2}M}$ that satisfies $l<r_m<r_0	$. So, we can write
\begin{equation}
r^*(r)=\int\frac{1}{1-\frac{r^2}{l^2}+\sqrt{\frac{r^4-r_m^4}{l^4}}}\ dr,
\end{equation}
and separate this way
\begin{equation}
r^*=\frac{r}{2}+\frac{r_0^2-l^2}{2}I_{11}+\frac{1}{2}I_{21}+\frac{r_0^2}{2}I_{22}+\frac{\left(r_0^2-l^2\right)^2}{2}I_{23},
\end{equation}
where
\begin{align}
I_{11}=&\ \int\frac{dr}{r^2-r_0^2}\quad,\quad I_{21}= \int\frac{r^2}{\sqrt{r^4-r_m^4}}\ dr,\\
I_{22}=&\ \int\frac{dr}{\sqrt{r^4-r_m^4}}\quad,\quad I_{23}=\int\frac{dr}{\bigl(r^2-r_0^2\bigr)\sqrt{r^4-r_m^4}}.\nonumber
\end{align}

The calculations gives this results
\begin{align}
I_{11}(r)=&\ \frac{1}{2r_0}\ln\left|\frac{(r_m+r_0)(r-r_0)}{(r_m-r_0)(r+r_0)}\right|\nonumber\\
I_{21}(r)=&\ r_m\left\{ \mathrm{F}\left(i\frac{r}{r_m},i\right)-\mathrm{E}\left(i\frac{r}{r_m},i\right)-\mathrm{F}\left(i,i\right)+\mathrm{E}\left(i,i\right)\right\} \\
I_{22}(r)=&\ \frac{1}{r_m}\left\{ \mathrm{F}\left(i\frac{r}{r_m},i\right)-\mathrm{F}\left(i,i\right)\right\}.\nonumber
\end{align}
Again, we have troubles with the computation of $I_{23}$ through incomplete elliptic integral of the third kind. We can use the same procedure from the preceding section. We separate the non finite part from the integrand,
\begin{equation}
\frac{1}{\bigl(r^2-r_0^2\bigr)\sqrt{r^4-r_m^4}}= \frac{1}{2r_0\sqrt{r_0^4-r_m^4}}\Biggl(\frac{1}{r-r_0}+\frac{2r_0\sqrt{r_0^4-r_m^4}-(r+r_0)\sqrt{r^4-r_m^4}}{(r^2-r_0^2)\sqrt{r^4-r_m^4}}\Biggr).
\end{equation}

Then, we integrate to obtain
\begin{equation}
I_{23}=\frac{1}{2r_0\sqrt{r_0^4-r_m^4}}\left(\ln\left|\frac{r-r_0}{r_0-r_m}\right|+\mathrm{Z}_{\alpha<0}(r)\right),
\end{equation}
where we define
\begin{equation}\label{zalm}
\mathrm{Z}_{\alpha<0}(r)=\int_{r_m}^r \frac{2r_0\sqrt{r_0^4-r_m^4}-(t+r_0)\sqrt{t^4-r_m^4}}{(t^2-r_0^2)\sqrt{t^4-r_m^4}}\ dt
\end{equation}
to be integrate through numerical methods.

Collecting all contributions and setting $r^*(r_m)=0$
\begin{align}
r^*=&\ \frac{r-r_m}{2}+\frac{\sqrt{r_0^4-r_m^4}}{4r_0}\left\{\ln\left(\frac{(r_0+r_m)(r-r_0)^2}{(r_0-r_m)^2(r+r_0)}\right)+Z_{\alpha<0}(r)\right\}\nonumber\\
&\ +\frac{r_m}{2}\left\{\mathrm F\left(i\frac{r}{r_m},i\right)-\mathrm E\left(i\frac{r}{r_m},i\right)-\mathrm F\left(i,i\right)+\mathrm E\left(i,i\right)\right\}\\
&\ +\frac{r_0^2}{2r_m}\left\{\mathrm F\left(i\frac{r}{r_m},i\right)-\mathrm F\left(i,i\right)\right\}. \nonumber
\end{align}

\subsection{Case $\alpha<0$ and $\frac{\kappa_{E}}{6\pi^2}M<l^2$}\label{int04}
The last integral is
\begin{equation}
r^*=\int\frac{dr}{ 1-\frac{r^2}{l^2}+\sqrt{\frac{r^4-r_m^4}{l^4}}},
\end{equation}
where $r_m=\sqrt[4]{\frac{\kappa_{E}}{6\pi^2}Ml^2}$. It is useful to define
\begin{equation}
r_0=\sqrt{\frac{r_m^4+l^4}{2l^2}}.
\end{equation}
Note that $r_m<r_0<l$. 

So, $r^*$ is given by
\begin{equation}
r^*=\frac{r}{2}-\frac{l^2-r_0^2}{2}I_{11}+\frac{1}{2}I_{21}+\frac{r_0^2}{2}I_{22}+\frac{\left(l^2-r_0^2\right)^2}{2}I_{23},
\end{equation}
where
\begin{align}
I_{11}=&\int\frac{dr}{r^2-r_0^2}\quad,\quad I_{21}=\int\frac{r^2}{\sqrt{r^4-r_m^4}}\ dr\\
I_{22}=&\int\frac{dr}{\sqrt{r^4-r_m^4}}\quad,\quad I_{23}=\int\frac{dr}{\bigl(r^2-r_0^2\bigr)\sqrt{r^4-r_m^4}}.\nonumber
\end{align}

The computation gives the following results
\begin{align}
I_{11}(r)=&\ \frac{1}{2r_0}\ln\left|\frac{(r_m+r_0)(r-r_0)}{(r_m-r_0)(r+r_0)}\right|\nonumber\\
I_{21}(r)=&\ r_m\left\{ \mathrm{F}\left(i\frac{r}{r_m},i\right)-\mathrm{E}\left(i\frac{r}{r_m},i\right)-\mathrm{F}\left(i,i\right)+\mathrm{E}\left(i,i\right)\right\} \\
I_{22}(r)=&\ \frac{1}{r_m}\left\{ \mathrm{F}\left(i\frac{r}{r_m},i\right)-\mathrm{F}\left(i,i\right)\right\}\nonumber\\
I_{23}(r)=&\ \frac{1}{2r_0\sqrt{r_0^4-r_m^4}}\left(\ln\left|\frac{r-r_0}{r_0-r_m}\right|+\mathrm{Z}_{\alpha<0}(r)\nonumber\right),
\end{align}
with $\mathrm{Z}_{\alpha<0}(r)$ given in eq. \ref{zalm}.

Collecting all contributions and setting $r^*(r_m)=0$
\begin{align}
r^*=&\ \frac{r-r_m}{2}+\frac{\sqrt{r_0^4-r_m^4}}{4r_0}\left\{\ln\left(\frac{r+r_0}{r_m+r_0}\right)+Z_{\alpha<0}(r)\right\}\nonumber\\
&\ +\frac{r_m}{2}\left\{\mathrm F\left(i\frac{r}{r_m},i\right)-\mathrm E\left(i\frac{r}{r_m},i\right)-\mathrm F\left(i,i\right)+\mathrm E\left(i,i\right)\right\}\\
&\ +\frac{r_0^2}{2r_m}\left\{\mathrm F\left(i\frac{r}{r_m},i\right)-\mathrm F\left(i,i\right)\right\}. \nonumber
\end{align}

\end{document}